\documentclass[floatfix,twocolumn]{revtex4-1}

\usepackage{graphicx} 
\usepackage{dcolumn} 
\usepackage{amsmath}
\usepackage{amssymb}
\usepackage{physics}
\usepackage{bm}
\usepackage{color}
\usepackage{times}
\usepackage{ulem}
\usepackage[justification=raggedright]{caption}

\usepackage{subcaption}
\usepackage{tabularx}
\usepackage{appendix}

\usepackage{amstext}
\usepackage{array}

\usepackage{diagbox}

\usepackage{bbm}

\definecolor{wine-stain}{rgb}{0.5,0,0} 
\definecolor{bblue}{rgb}{0,0.0,0.5} 

\usepackage[colorlinks=true
,urlcolor=bblue
,anchorcolor=blue
,citecolor=wine-stain
,filecolor=blue
,linkcolor=blue
,menucolor=blue
,pagecolor=blue
,linktocpage=true
,pdfproducer=medialab
,linktoc=all 
]{hyperref}

\usepackage{multirow}

\usepackage{booktabs}  

\allowdisplaybreaks 

\usepackage{setspace}
\usepackage{custom_commands}

\newcommand{\es}{\text{\scriptsize{(ES)}}}
\newcommand{\fs}{\text{\scriptsize{(FS)}}}
\newcommand{\bcs}{\text{\scriptsize{(BCS)}}}
\newcommand{\dw}{\text{\scriptsize{(DW)}}}
\newcommand{\wV}[1]{\wtil{V}_{#1;\text{\scriptsize{J}}}}
\newcommand{\wU}[1]{\wtil{U}_{#1;\text{\scriptsize{J}}}}
\newcommand{\wUe}[1]{\wtil{U}_{#1;\text{\scriptsize{J}}}^\es}
\newcommand{\wUf}[1]{\wtil{U}_{#1;\text{\scriptsize{J}}}^\fs}
\newcommand{\V}[1]{V_{#1;\text{\scriptsize{J}}}}
\newcommand{\U}[1]{U_{#1;\text{\scriptsize{J}}}}
\newcommand{\f}[1]{f_{#1;\text{\scriptsize{J}}}}

\newcommand{\D}[1]{\Dl_{#1;\text{\scriptsize{J}}}}
\newcommand{\ph}[1]{\Phi_{#1;\text{\scriptsize{J}}}}

\makeatletter
\def\l@subsection#1#2{}
\def\l@subsubsection#1#2{}
\makeatother

\begin{document}

\title{
Instabilities of Weyl-loop semi-metals
}

\date{\today}

\author{Shouvik Sur$^{1}$ and Rahul Nandkishore$^{2}$}
\affiliation{
$^1$National High Magnetic Field Laboratory,\\
Florida State University, Tallahassee, Florida 32306, USA \\
$^2$Department of Physics and Center for Theory of Quantum Materials,\\
Univeristy of Colorado, Boulder, Colorado 80309, USA
}

\begin{abstract}
We study Weyl-loop semi-metals with short range interactions, focusing on the possible interaction driven instabilities. We introduce an $\epsilon$ expansion regularization scheme by means of which the possible instabilities may be investigated in an unbiased manner through a controlled weak coupling renormalization group calculation. The problem has enough structure that a `functional' renormalization group calculation (necessary for an extended Fermi surface) can be carried out analytically. The leading instabilities are identified, and when there are competing degenerate instabilities a Landau-Ginzburg calculation is performed to determine the most likely phase.  In the particle-particle channel, the leading instability is found to be to a fully gapped chiral superconducting phase which spontaneously breaks time reversal symmetry, in agreement with general symmetry arguments suggesting that Weyl loops should provide natural platforms for such exotic forms of superconductivity. In the particle hole channel, there are two potential instabilities - to a  gapless Pomeranchuk phase which spontaneously breaks rotation symmetry, or to a fully gapped insulating phase which spontaneously breaks mirror symmetry. The dominant instability in the particle hole channel depends on the specific values of microscopic interaction parameters. 
\end{abstract}

\maketitle

\hypersetup{linkcolor=bblue}
\tableofcontents



\section{Introduction} \label{sec:intro}
The most generic metallic states occur in systems that host Fermi surfaces whose dimension is one less than the dimension of the system. 
In the presence of effective short-range interactions among the fermions, these metallic states are  described by the Tomonaga-Luttinger model in one dimension \cite{Tomonaga1950,Luttinger1963, Mattis1965, Dzyaloshinski1974, Haldane1981a, Haldane1981b, Giamarchi-Bk}, and frequently by Landau's Fermi liquid theory above one dimension \cite{Landau1957}.
Comparatively less common metallic states are realized in systems where a  filled valence band touches a conduction band.
These \textit{semi-metallic} states  possess gapless excitations about a zero-energy manifold with dimension   two or more below the spatial dimension of the system. 
Although semi-metals have been theoretically investigated since at  least 1970s \cite{Abrikosov1970}, their properties have garnered considerable interest in the last two decades with the advent of graphene \cite{g1, g2, graphenereview} and other varieties of Dirac materials \cite{ti1, ti2, ti3, ti4, ti5, ti6, ti7, tireview1, tireview2, tireview3, weyl1, weyl2, weyl3, weylreview1, weylreview2, weylreview3}. Most of the known semi-metals contain a discrete set of gapless points in the bulk.
However, in recent years, three dimensional semi-metals with a ring of gapless points have become a possibility \cite{Burkov2011, loop1, Bian2016, loop2, loop3, loop4, loop5, loop6, loop7, loop8, loop9}. Theoretical investigation into the effect of the weakly screened long-range Coulomb interaction on these  \textit{Weyl-loop} semi-metals suggests that single-particle excitations survive at low energy, as quantum fluctuations render the Coulomb interaction marginally irrelevant \cite{Huh2015}.
However, 
strong short-range interactions can lead to symmetry-breaking instabilities. Indeed, it has been argued that such `Weyl loop' systems may serve as ideal playgrounds for realizing exotic forms of superconductivity \cite{vik1, Nandkishore2016}. 
However a systematic and unbiased treatment of the potential interaction driven instabilities of Weyl loop systems remains to be performed.



In this paper we investigate the effect of short-range interactions on a Weyl-loop semi-metal, and identify the symmetry broken states that are probable at finite interactions.

The paper is organized as follows.
In section \ref{sec:model} we introduce the continuum model  whose low energy properties are the subject of this work.
A generalization based on tuning the dispersion of the fermions is developed, which enables access to finite coupling instabilities within the regime of applicability of a controlled weak coupling perturbation theory. 
In sections \ref{sec:RG} to \ref{sec:non-BCS} the low energy properties of the  perturbatively accessible sector of the generalized model is analyzed within a renormalization group (RG) scheme based on mode elimination. The RG is shown to have enough structure that a `functional' analysis (necessary for an extended Fermi surface) can be carried out analytically. The particle-particle and particle-hole channels are found to decouple. 
We deduce the fixed points of the running couplings in  the particle-particle and particle-hole channels respectively. 
In section \ref{sec:suscep} the most likely instability is identified through an analysis of  anomalous dimensions of the susceptibilities of various pairing channels, combined with a Landau-Ginzburg analysis. In the particle particle channel we find an instability to a novel form of superconductivity, wherein the order parameter is fully gapped and chiral, and spontaneously breaks time reversal symmetry. In the particle-hole channel there are two potential instabilities (with the dominant one being determined by microscopic values of interaction parameters): either a Pomeranchuk instability to a gapless phase that spontaneously breaks rotation symmetry, or an excitonic instability to a gapped (trivial) insulating state that spontaneously breaks mirror symmetry. 
Finally, in section \ref{sec:conclusion} we conclude with a discussion of our results.


\section{Model} \label{sec:model}
In this section we derive an effective theory that is appropriate for understanding the universal low energy properties of the Weyl-loop semi-metal in the presence of short range interactions.
Since short range interactions are expected to be strongly irrelevant in the presence of linear band-touching, we develop a convenient generalization of the model in terms of the degree of  band-curvature, which allows us to access interaction driven instabilities within the regime of applicability of a weak coupling RG. 

\subsection{Non-interacting theory} \label{sec:S-0}
The simplest realistic description of non-interacting fermions whose dispersion admits a nodal line Fermi surface in three dimensions is given by \cite{Huh2015,Nandkishore2016}
\begin{align}
 S_0 = \int dK ~ \Psi^{\dagger}(K) \lt[ ik_0 \sig_0 + E(\mbf K) \rt] \Psi(K),
\label{eq:S_0}
\end{align}
where $dK \equiv \frac{dk_0 dK_x dK_y dK_z}{(2\pi)^4}$, $k_0$ is the Euclidean (Matsubara)  frequency, $\mbf K$ denotes three dimensional momentum, $\sig_0$ is the $2\times 2$ identity matrix, and $\Psi(K) = \trans{(c_1(K), \quad c_2(K))}$ is a spinor  representating fermions $(c_{1,2}(K))$ from $1,  2$ orbitals.
The dispersion is
\begin{align}
E(\mbf K) 
& = \wtil v_r(|\vec K|) ~ (|\vec K| - \kap)~ \sig_1 + v_z K_z \sig_2,
\label{eq:disp}
\end{align}
where $\wtil v_r(|\vec K|) = (|\vec K| + \kap)/(2m)$.
Here we have distinguished between the three dimensional momentum from its projection, $\vec K$, on the plane of the Weyl-loop. 
We have chosen the loop to lie on the $x-y$ plane, and it is defined by $|\vec K| = \sqrt{K_x^2 + K_y^2} = \kap$.
Here $\sig_1$ and $\sig_2$ are the first two Pauli matrices which encode the orbital degrees of freedom, and $m$ and $v_z$ are bandstructure parameters.
We note that at finite doping, i.e. away from perfect compensation, the non-interacting theory is modified by replacing $ik_0 \mapsto ik_0 - \mu$. Our theory will focus on $\mu =0$. 

\begin{figure}[!t]
\centering
\includegraphics[width=0.9\columnwidth]{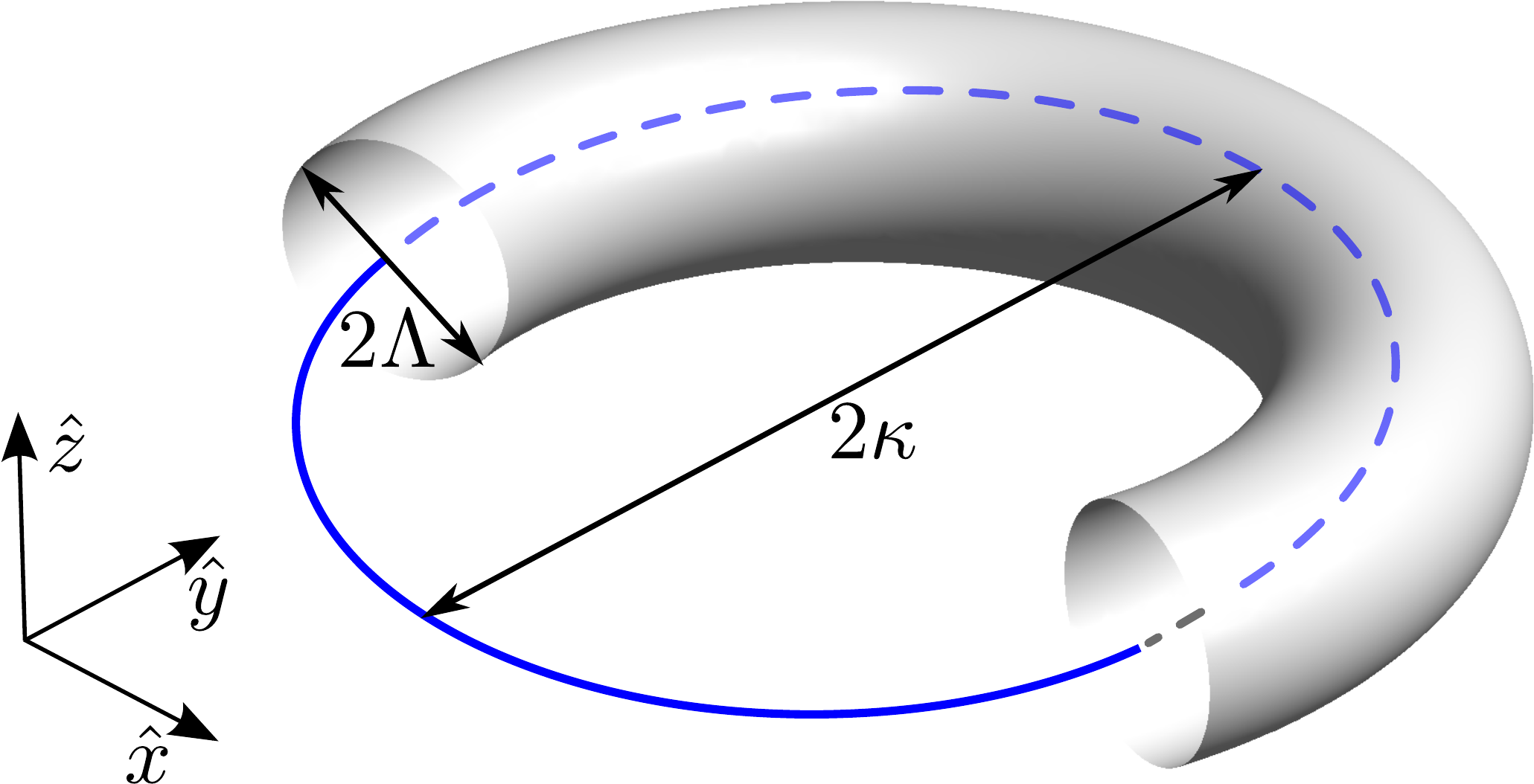}
\caption{The (blue) ring on the $x-y$ plane is the nodal loop. The torus enveloping the loop is the UV cutoff.}
\label{fig:torus}
\end{figure}

Diagonalizing $E(\mbf K)$ yields two bands that disperse as
\begin{align}
\xi_\pm(\mbf K) = \pm \sqrt{\wtil v_r^2({\vec K}) (|\vec K| - \kap)^2 + v_z^2 K_z^2}.
\end{align}
Since the chemical potential $\mu = 0$, the ground state is defined by the configuration where $|\vec K| = \kap$ and $K_z = 0$, which precisely corresponds to the loop.
Thus at low energy $|\vec K| \approx \kap$ and $K_z \approx 0$,  and the dispersion, \eq{eq:disp}, can be approximated to
\begin{align}
\veps(k_r, k_z) = v_r ~k_r~ \sig_1 + v_z k_z \sig_2,
\end{align}
where $v_r \equiv \wtil v_r(\kap) = \kap/m$, and $k_r \equiv |\vec K| - \kap$ and $k_z$ are deviations of momentum in the radial and $z$ directions, respectively.
The band dispersion simplifies to $\xi_\pm(k_r, k_z) = \pm \sqrt{v_r^2 k_r^2 + v_z^2 k_z^2}$.

We scale $(k_0, k_z) \mapsto v_r(k_0, k_z/v_z)$, and identify the long wavelength fluctuations of the fermions (low energy modes) through the relation, 
\begin{align}
\Psi(\tau, \mbf r) \sim e^{i\vec \kap \cdot \mbf r} ~\psi(\tau,\mbf r) + \mbox{ fast modes},
\end{align}
where $\vec \kap = \kap(\cos \theta, \sin \theta,0)$ with $\theta$ specifying the position on the loop.
We further sharpen the definition of the low energy modes by requiring that the momentum carried by these modes  to be such that  $\sqrt{k_r^2 + k_z^2} < \Lam \ll \kap$, where $\Lam$ is a UV cutoff on the $k_r \textrm{-} k_z$ plane, measured from the loop (see \fig{fig:torus}).
Integrating out the modes which modulate over length scales $\lesssim \Lam^{-1}$  we obtain 
\begin{align}
 S_0 &\simeq \frac{v_r^3 \kap}{v_z} \int_0^{2\pi} \frac{d\theta}{2\pi} \int_{-\infty}^{\infty}  \frac{dk_0 dk_r dk_z}{(2\pi)^3} ~ f_\Lam(k_r, k_z) \nn \\
& \qquad \times \psi^{\dagger}(k_0,\vec k, \theta) \lt[ ik_0 \sig_0 + k_r \sig_1 +  k_z \sig_2 \rt] \psi(k_0, \vec k, \theta) ,
\label{eq:S_0-2}
\end{align}
where $f_\Lam(k_r, k_z)$ is a cutoff function which suppresses  modes with $k_r^2 + k_z^2 \gtrsim \Lam^2$ \cite{cutoff}.
We choose $f_\Lam(k_r, k_z)$ to be rotationally symmetric in $k_r \textrm{-} k_z$ plane.
The two dimensional vector $\vec k$ is the deviation of momentum from the loop, and is defined on the $k_r \textrm{-} k_z$ plane  by $\vec k \equiv (k_r, k_z)$.
We emphasize that $\vec k$ is not linearly related to the deviations from the loop in  $\mbf K$ space.
In particular, under inversion $\mbf K \mapsto - \mbf K$, but  $\vec k = (k_r, k_z) \mapsto  (k_r, -k_z)$.

\subsubsection{Symmetries} \label{sec:symmetries}
\begin{table*}[!t]
\centering
\begin{tabular}{l | l }
Symmetry & Operation \\
\hline \\[-2ex]
$\theta$-rotation & $\mc{R}_\theta : \theta \mapsto \theta + \theta_0$  \\[0.5ex]
Pseudospin-$U(1)_\infty$ & $\mc{R}_\vphi : \psi \mapsto \exp(i \xi_\theta \mc{J}) ~\psi$  \\[0.5ex]
Charge-$U(1)_\infty$ & $\psi \mapsto \exp(i \om_\theta) ~\psi$  \\[0.5ex]
Mirror plane ($\mc P_z$) & $k_z \mapsto - k_z$ and $\psi \mapsto \sig_1 ~\psi$ \\[0.5ex]
Anti-unitary ($\mc P_\mbf{K}$) 
& $\mbf K \mapsto - \mbf K$ 
and $\{\psi, \psi^\dag \} \mapsto \{ \trans{(\psi^\dag)}, - \trans{\psi} \}$
 \\[0.5ex]
Anti-unitary  ($\mc P_z \mc P_\mbf{K}$) 
& $\mbf K \mapsto - \mbf K$, $\{\psi, \psi^\dag \} \mapsto \{ \sig_1  \trans{(\psi^\dag)}, - \trans{\psi} \sig_1 \}$ \\[0.5ex]
Antiunitary  ($\mc P_0$) 
& $k_0 \mapsto - k_0$, $\{\psi, \psi^\dag \} \mapsto \{ \sig_2 \trans{(\psi^\dag)},  \trans{\psi} \sig_2\}$
 \\[0.5ex]
Antiunitary ($\mc P_z \mc P_0$) 
& $k_0 \mapsto - k_0$ $\{\psi, \psi^\dag \} \mapsto \{ \sig_3  \trans{(\psi^\dag)},  \trans{\psi} \sig_3 \}$
\end{tabular}
\caption{The symmetries of the Gaussian fixed point action \eq{eq:S_0-2}. $\theta_0$ is a constant, and $\om_\theta$ and $\xi_\theta$ are functions of $\theta$. The first three are continuous symmetries, while the last five are discrete. The first three antiunitaries square to $-1$ while the last squares to $+1$.
}
\label{tab:symmetries}
\end{table*}
The dynamics of the non-interacting fermions described by \eq{eq:S_0-2} enjoys a set of continuous and discrete symmetries.
In this section we describe these symmetries, and the respective symmetry transformations.

Since the fermion dynamics is  independent of the position of the fermionic momentum on the loop, the loop coordinate $\theta$ acts as a label for the $\psi$ fields.
The cyclic nature of $\theta$ leads to three continuous  symmetries of $S_0$.
The first is a $SO(2)$ rotational invariance of the action under  $\mc{R}_\theta : \theta \mapsto \theta + \theta_0$.
In order to isolate the second symmetry let us write the spatial part of the propagator in terms of polar coordinates $(k,\vphi)$ as 
$k (\cos\vphi ~\sig_1 + \sin\vphi ~ \sig_2)$, where $(k_r, k_z) = (k \cos\vphi, k \sin\vphi)$.
Under the transformation $\mc{R}_\vphi: \psi \mapsto e^{i\xi_\theta \mc{J}} \psi$, with $\mc{J} = \sig_3/2 - i\dow_\vphi$,  the lagrangian 
\begin{eqnarray}
L_0[k_0, k, \theta, \vphi; \psi] &=& \psi^{\dagger}(k_0, k, \theta, \vphi) 
[ ik_0 \sig_0\\ &+& k (\cos\vphi ~\sig_1 + \sin\vphi ~ \sig_2)]  \nonumber
\psi(k_0, k, \theta, \vphi)
\end{eqnarray}
transforms to $L_0[k_0, k, \theta, \vphi + \xi_\theta; \psi]$, implying $\mc{R}_\vphi$ is a symmetry of $S_0$.
$\mc{R}_\vphi$ corresponds to a  rotation in the plane perpendicular to the loop at each point $\theta$.
The third symmetry is the invariance of the action under a $\theta$-dependent $U(1)$  transformation $\psi \mapsto e^{i \omega_\theta} \psi$.
Since the latter two symmetry transformations are locally defined on the loop,  they lead to distinct emergent $U(1)_\infty$ symmetries which we will distinguish as pseudospin-$U(1)_\infty$ and charge-$U(1)_\infty$, respectively.
While the former corresponds to the conservation of $\hat \theta$ component of total angular momentum, the latter originates from particle number conservation at each $\theta$.
We note that the charge-$U(1)_\infty$ symmetry is present in any non-interacting theory where the single particle dispersion is minimized on a degenerate manifold.
Since short-range interactions mix momenta at  different parts of the loop, these $U(1)_\infty$ symmetries are broken by generic scatterings among the fermions.
Nevertheless, it is possible for  subgroups of the $U(1)_\infty$ symmetries to emerge at fixed points of the interacting theory \cite{Shankar1994}.

The action is also invariant under three sets of discrete  transformations.
The first is a mirror-plane symmetry which originates from the symmetry between the dynamics above and below the $k_z = 0$ plane. 
It is effected by the transformation $\mc{P}_z^{-1} ~L_0[k_0, k, \theta, \vphi; \psi] ~\mc{P}_z = L_0[k_0, k, \theta, -\vphi; \psi]$, where the `operator' $\mc{P}_z$ flips the sign of $k_z$ such  that $\mc{P}_z^{-1} ~ \{k_0, k, \theta, \vphi \} ~\mc{P}_z = \{k_0, k, \theta, -\vphi \}$ and $\mc{P}_z^{-1} \psi(k_0, k, \theta, \vphi) \mc{P}_z = \sig_1  \psi(k_0, k, \theta, -\vphi)$.
The second is a pair of  `anti-unitary' symmetries, the first of which is defined through the transformation  $\mc{P}_{\mbf{K}}^{-1} ~L_0[k_0, k, \theta, \vphi; \psi] ~\mc{P}_\mbf{K} = L_0[k_0, k, \theta + \pi, -\vphi; \psi]$.
Here $\mc{P}_\mbf{K}$ inverts the three-momentum $\mbf K$, and acts on the fermion fields as $\mc{P}_{\mbf{K}}^{-1} \{\psi(k_0, k, \theta, \vphi), \psi^\dag(k_0, k, \theta, \vphi) \} \mc{P}_{\mbf{K}} = \{\psi^*(k_0, k, \theta + \pi, -\vphi), - \trans{\psi}(k_0, k, \theta + \pi, -\vphi)\}$ with $\psi^* \equiv \trans{(\psi^\dag)}$.
The second element of the pair is obtained by combining $\mc P_z$ with $\mc P_{\mbf K}$. We note that while these are symmetries of the action, they act on the {\it Hamiltonian } in an unusual way. In particular at the level of a first quantized Hamiltonian they change the sign of the $\sigma_1$ term, and hence effectively connect the Hamiltonian with $|\vec{K}| - \kappa = k_r < 0$ to the Hamiltonian with $ |\vec{K}| - \kappa > 0$.
The last of the sets of discrete symmetries is another pair of antiunitary transformations whose first element is defined by  $\mc{P}_{0}^{-1} ~L_0[k_0, k, \theta, \vphi; \psi] ~\mc{P}_0 = L_0[-k_0, k, \theta, \vphi; \psi]$, where $\mc P_0$ inverts the Euclidean frequency $k_0$ and transforms the fields as
$\mc{P}_{0}^{-1} \{\psi(k_0, k, \theta, \vphi), \psi^\dag(k_0, k, \theta, \vphi) \} \mc{P}_{0} = \{\sig_2 \psi^*(-k_0, k, \theta, \vphi), \trans{\psi}(-k_0, k, \theta , \vphi) \sig_2 \}$.
The second element of the pair is obtained by combining $\mc P_z$ with $\mc P_0$. 
We summarize all the symmetries of $S_0$ in Table \ref{tab:symmetries}.

\subsubsection{Generalization} \label{sec:generalization}
As we will show later in the paper, in order to obtain a controlled truncation of the beta functions  of various operators  it is convenient to generalize the linear band-touching model to higher order band-touchings.
We use the polar coordinates introduced in section \ref{sec:symmetries} to generalize the dispersion of fermions as  $\veps_\eta(k,\vphi) = k^{\eta} ~ (\cos\vphi ~ \sig_1 + \sin\vphi ~\sig_2)$ for any real number $\eta >0$.
The band dispersion is modified accordingly as $\xi_{\eta;\pm} (k, \vphi) = \pm k^{\eta}$. 
Thus, the dynamics of the  non-interacting fermions with $\eta$-th order band-touching is given by 
\begin{widetext}
\begin{equation}
 S_{\eta;0} = \frac{v_r^3 \kap }{v_z} \int_{-\pi}^{\pi} \frac{d\theta d\vphi}{(2\pi)^2} \int_{-\infty}^{\infty} \frac{dk_0 }{2\pi}\int_{0}^{\infty} \frac{dk ~ k}{2\pi}  ~ f_\Lam(k)
 \psi^{\dagger}(k_0,k, \theta, \vphi) \lt[ ik_0 \sig_0 + k^{\eta} ~ ( \cos{(\vphi)} \sig_1 + \sin{(\vphi)} \sig_2 ) \rt] \psi(k_0,k, \theta, \vphi).
\label{eq:S_n0-1}
\end{equation}
\end{widetext}
Since the value of $\eta$ does not affect the symmetry transformations, $S_{\eta;0}$ and $S_{1;0}$ share the same set of symmetries.
The Gaussian fixed point described  by $ S_{\eta;0}$ is invariant under the following choice of scaling, 
\begin{align}
[\theta], [\vphi] = 0, \quad [k] = 1, \quad [k_0] = \eta, \quad [\psi] = - (\eta + 1),
\label{eq:scaling-1}
\end{align}
where the quantity $X$ scales as $X' = e^{[X] d\ell} X$ with $d\ell$ being a  logarithmic \textit{energy} scale.
We note that within our scheme the radius of the loop ($\kap$) is dimensionless, which implies coarse-graining towards the loop in momentum space \cite{Polchinski1992,Shankar1994}.

\subsection{Interactions} \label{sec:S-int}
In this section 
we introduce the vertices describing  instantaneous short-range interactions that are consistent with the discrete symmetries of the non-interacting theory.
After scaling $k_0$ and $K_z$, as was done for $S_0$, to obtain 
\begin{widetext}
\begin{equation}
\wtil S_{int} = \lt(\frac{v_r^2}{v_z} \rt)^3 \sum_{\mu,\nu=0}^{3} \int_{\Lam} \lt(\prod_{n=1}^4 dK_n \rt) (2\pi)^4 ~ \dl^{(4)}(K_1 - K_2 + K_3 - K_4)  
\mc{U}_{\mu,\nu}(\{\mbf K_n \}) ~  \lt[ \psi^{\dagger}(K_1) \sig_\mu \psi(K_2) \rt] \lt[ \psi^{\dagger}(K_3) \sig_\nu \psi(K_4) \rt], 
\label{eq:S-int-1}
\end{equation}
\end{widetext}
where $\int_\Lam dK = \int dK ~ f_\Lam(|\vec K|-\kap, K_z)$.
We assume $\mc{U}_{\mu,\nu}(\{\mbf K_n \}) = \mc{U}_{\nu,\mu}(\{\mbf K_n \})$ to be real valued functions of momentum.
In general, there are $16$ vertices corresponding to four choices each for $\mu$ and $\nu$.
In this paper we focus on only those vertices that are invariant under the discrete symmetries of $S_0$, viz. $\mc P_z$, $\mc P_{\mbf K}$, and $\mc P_0$.
The terms that have odd parity  under  these symmetries are marked with  corresponding labels in Table \ref{tab:prohibited}.
\begin{table}[!t]
\centering
\begin{tabular}{| c || c | c | c | c |}
\hline
\diagbox{$\nu$}{$\mu$} & $\quad 0 \quad$ & $\quad 1 \quad$ & $\quad 2 \quad$ & $\quad 3 \quad$ \\
\hline \hline
$0$ & $ ~ $ & $\mc P_0$ & $\mc P_z, \mc P_{\mbf K}, \mc P_0$ & $\mc P_z$, $\mc P_0$ \\
\hline
$1$ & $\mc P_0$ & $ ~ $ & $\mc P_z, \mc P_{\mbf K} $ & $\mc P_z$ \\
\hline
$2$ & $\mc P_z, \mc P_{\mbf K}, \mc P_0$ & $\mc P_z, \mc P_{\mbf K}$ & $ ~ $ & $\mc P_{\mbf K}$ \\
\hline
$3$ & $\mc P_z, \mc P_0$ & $\mc P_z$ & $\mc P_{\mbf K}$ & $ ~ $ \\
\hline
\end{tabular}
\caption{The vertices in \eq{eq:S-int-1} that are disallowed by the discrete symmetries of $S_0$ (see Table \ref{tab:symmetries}). 
The $(\mu, \nu)$-th cell in the table represents the corresponding vertex. 
The cells are labeled by the symmetry transformations under which the parity of the vertex is odd.}
\label{tab:prohibited}
\end{table}
From the table we see that only those terms for which $\mu = \nu$ have the same parity as $S_0$.
Thus, the minimal set of interactions that respect the discrete symmetries of $S_{\eta;0}$ is given by
\begin{align}
& S_{int} = \frac{v_r^6}{v_z^3} \int_\Lam \lt(\prod_{n=1}^4 dK_n \rt) ~ (2\pi)^4 \dl^{(4)}(K_1 - K_2 + K_3 - K_4) \nn \\
& \times \sum_\mu \wtil g_{\mu}(\{\what K_n\}) \lt( \psi^{\dagger}(K_1) ~\sig_\mu~ \psi(K_2) \rt) \lt( \psi^{\dagger}(K_3) ~\sig_\mu~ \psi(K_4) \rt),
\label{eq:S_int-2}
\end{align}
where $\what K_n = \vec K_n /|\vec K_n|$, and 
we have Taylor expanded $\mc U_{\mu,\mu}(\mbf K_1, \mbf K_2, \mbf K_3, \mbf K_4)$ about the loop to obtain the coupling functions $\wtil g_\mu(\what K_1, \what K_2, \what K_3, \what K_4)$.
Since, $\what K$ and $\theta$ are physically equivalent, $\wtil g_\mu(\{\what K_n\})$ are  functions of the loop coordinates only.

We add \eq{eq:S_n0-1} to  \eq{eq:S_int-2}, and scale $\psi \mapsto \sqrt{(v_z/v_r^3)} ~ \psi$, to obtain the action for the interacting theory for any $\eta$.
It is useful to introduce  $(\gamma_0,\gamma_1, \gam_2, \gam_3)  \equiv (\sig_3, \sig_2, - \sig_1, \sig_0)$, and define the conjugate field
\begin{align}
\bar \psi(K) = \psi^{\dag}(K) ~\gam_0,
\label{eq: conjugate}
\end{align}
such that the generalized low energy effective  theory is 
\begin{widetext}
\begin{align}
S(\eta) &= i\int_\Lam dK  \bar\psi(K) \lt[ k_0 \gam_0 + ((|\vec K| - \kap)^2 + K_z^2)^{\frac{(\eta-1)}{2}} ~ ((|\vec K| - \kap)~ \gam_1 +  K_z \gam_2) \rt] \psi(K) \nn \\
&+ \int_\Lam \lt(\prod_{n=1}^4 dK_n \rt) ~(2\pi)^4 ~\dl^{(4)}(K_1 - K_2 + K_3 - K_4) 
 \Bigl[ \frac{g_{1}(\{\what K_n\})}{\kap} \lt( \bar \psi(K_1) \gam_0  \psi(K_2) \rt) \lt( \bar \psi(K_3) \gam_0 \psi(K_4) \rt) \nn \\
&- \frac{g_{2+}(\{\what K_n\})}{\kap} \Bigl\{\lt( \bar \psi(K_1) \gam_1 \psi(K_2) \rt) \lt( \bar \psi(K_3) \gam_1 \psi(K_4) \rt) + (\gam_1 \rtarw \gam_2) \Bigr\}  \\
&- \frac{g_{2-}(\{\what K_n\})}{\kap} \Bigl\{\lt( \bar \psi(K_1) \gam_1 \psi(K_2) \rt) \lt( \bar \psi(K_3) \gam_1 \psi(K_4) \rt) - (\gam_1 \rtarw \gam_2) \Bigr\}
+ \frac{g_{3}(\{\what K_n\})}{\kap} \lt( \bar\psi(K_1)  \psi(K_2) \rt) \lt( \bar\psi(K_3) \psi(K_4) \rt) \Bigr],\nn
\label{eq:action}
\end{align}
\end{widetext}
where $(g_1, g_{2+}, g_{2-}, g_3) \equiv \frac{\kap}{v_z} ~ (\wtil g_0, \half(\wtil g_1 + \wtil g_2), \half(\wtil g_1 - \wtil g_2), \wtil g_3)$.
We note that in contrast to the Gaussian part, the interaction terms generally do not admit a straightforward decomposition in terms of angular patches because the short-range scatterings mix angular coordinates.
Nevertheless, the diagonal structure of the Gaussian part in terms of the patch index $\theta$ is useful for the evaluation of quantum corrections.

From \eq{eq:action} we deduce the bare propagator
\begin{align}
\avg{ \psi_{a}(K') ~ \bar \psi_{b}(K) }_0 
& = (2\pi)^4 ~ \dl^{(4)}(K - K') ~ \mc G_{a,b}(K),
\end{align}
where
\begin{align}
\mc G(K) \equiv \mc G(k_0, k, \vphi) = -i~ \frac{k_0 \gamma_0 +  k^{\eta}  \lt(\cos{\vphi} ~\gam_1 + \sin{\vphi} ~\gam_2 \rt)}{k_0^2 + k^{2\eta}}
\label{eq: propagator}
\end{align}
By applying the scaling relations in \eq{eq:scaling-1} to the interaction vertices, we obtain the scaling dimension for the coupling functions,
\begin{align}
 [g_i] = \eta - 2.
\label{eq:scaling-2}
\end{align}
Therefore, for $\eta < 2$ ($\eta>2$) the interactions are irrelevant (relevant) at the Gaussian fixed point.
In the presence of irrelevant interactions the Gaussian fixed point is stable and has a finite basin of attraction, whose volume in coupling space is controlled by the parameter $\eps = 2 - \eta$.
Thus, it is expected that for $\eps>0$ weak short range interactions cannot lead to new phases, and the nodal loop  semi-metal is stable.
The Weyl-loop semi-metal corresponds to $\eps=1$, where the short-range interactions have bare scaling dimension $-1$, and are strongly irrelevant.
Although irrelevant in RG sense, microscopically strong interactions can still drive the system towards a non-trivial phase by pushing the couplings out of  the basin of attraction of the Gaussian fixed point.
Such finite-coupling instabilities of the $\eta=1$ ($\eps = 1$) system is the  focus of this paper.
However, at $\eta=1$ the bare scaling dimension of the couplings are $\ordr{1}$, which obstructs a controlled access to the potential finite-coupling fixed points and instabilities.
In order to achieve perturbative control  we turn to the limit where $0 \leq \eps \ll 1$.
In particular, $\eps = 0$ corresponds to a Weyl-loop  semi-metal with quadratic band-touching.
Here the short-range interactions are either marginally relevant or irrelevant, as is the case for  Fermi surfaces with unit codimension.
Motivated by the smooth interpolation between the quadratic and linear band-touching models at tree level, we   analyze the finite coupling instabilities close to $\eta=2$, using $\epsilon$ as the control parameter. 
In the spirit of all $\eps$ expansions, it is hoped that the small $\eps$ analysis would be able to  access qualitative  elements of the $\eta=1$ theory \cite{nonlocality}.
We note that our approach is complementary to reference \cite{Senthil2009}, where the codimension of a one-dimensional Fermi surface was used as a tuning parameter to controllably access a finite coupling pairing instability.
Similar strategies based on tuning the dispersion of the dynamical modes  \cite{Nayak1994,Mross2010}, and the codimension of the Fermi surface  \cite{Dalidovich2013,Sur2014,Mandal2015,Sur2016} have been applied to the study of strongly coupled field theories in the presence of a Fermi surface.



\section{Renormalization group} \label{sec:RG}
\begin{figure}[!t]
\centering
\includegraphics[width=0.9\columnwidth]{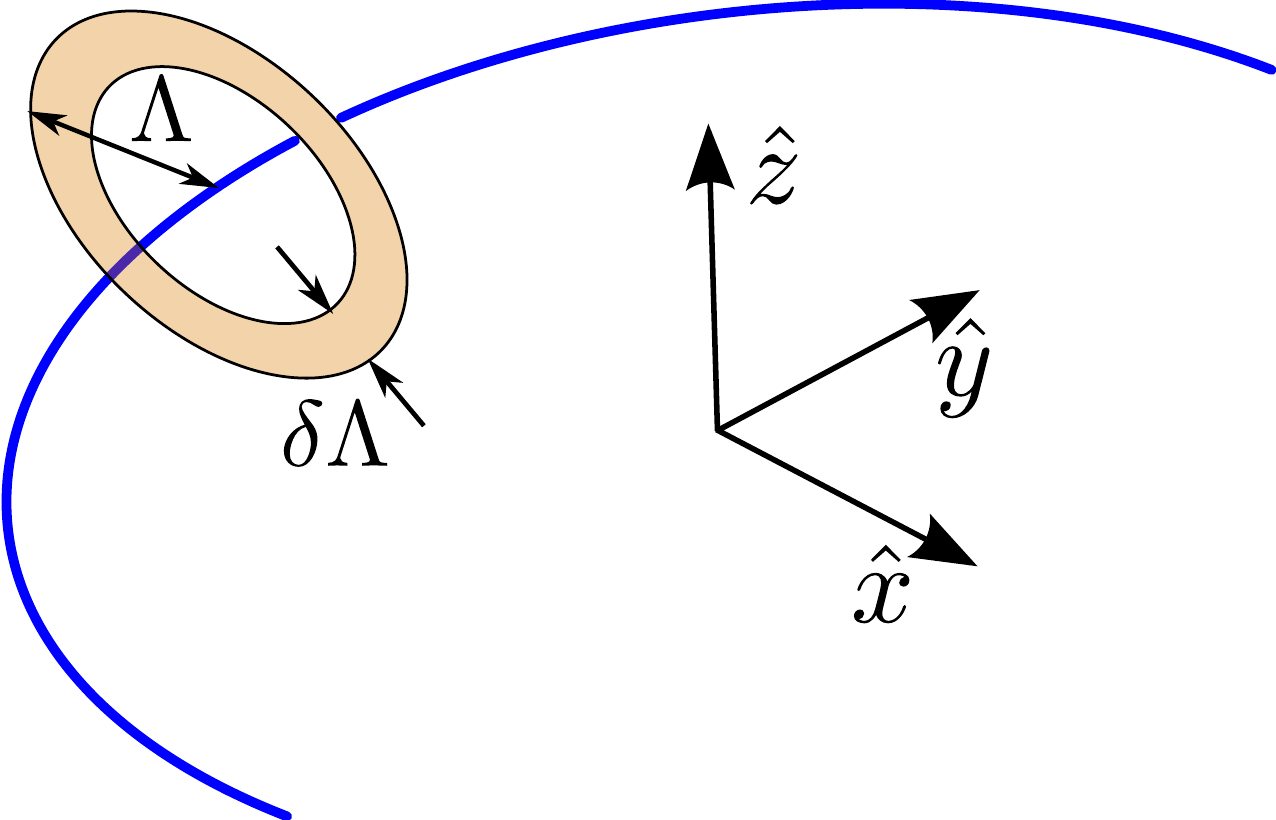}
\caption{Schematic of the strategy for coarse-graining towards the loop. Here we show a segment of the loop (blue arc). At each point of the loop we coarse-grain in the $k_r-k_z$ plane (shaded annular disk).}
\label{fig:coarse-grain}
\end{figure}
\begin{figure}[!]
\centering
\begin{subfigure}{0.45\columnwidth}
\includegraphics[width=0.95\columnwidth]{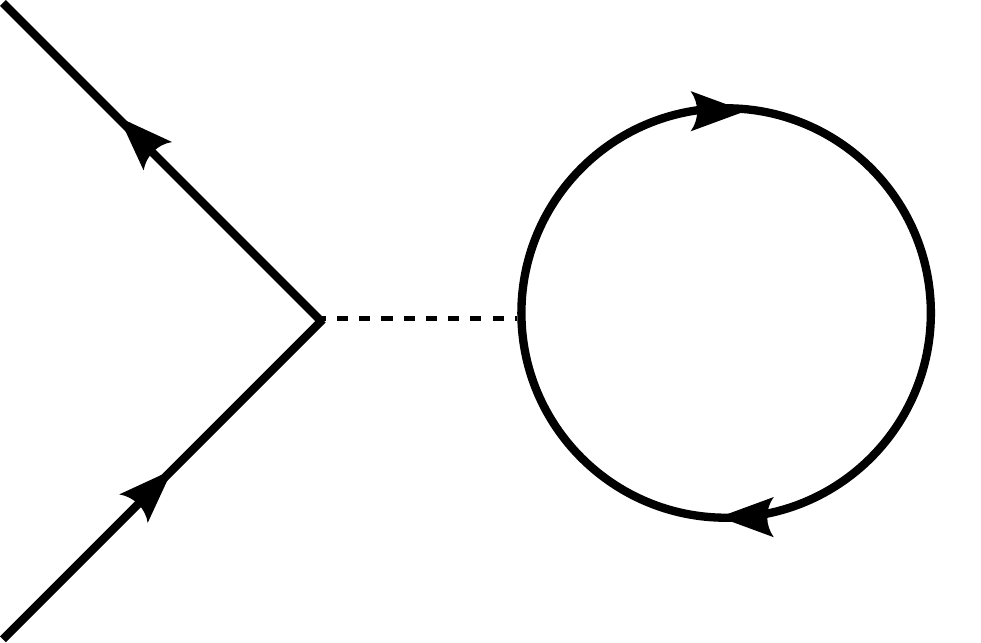}
\caption{Fock diagram}
\label{fig:fock}
\end{subfigure}
\hfill
\begin{subfigure}{0.45\columnwidth}
\includegraphics[width=0.95\columnwidth]{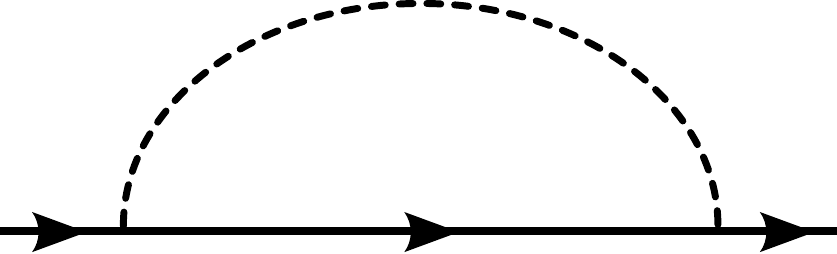}
\caption{Hartree diagram}
\label{fig:hartree}
\end{subfigure}
\caption{The Hartree-Fock diagrams that renormalize the self-energy at one-loop. Here they vanish identically.}
\label{fig:HF}
\end{figure}

In this section we outline our RG scheme for understanding the low energy properties of the Weyl-loop semi-metal in the presence of short-range interactions.
We will use the Wilsonian RG scheme due to Shankar \cite{Shankar1994} to derive  the beta functions. 
In particular, we coarse-grain towards the nodal loop by eliminating modes that lie in the region $[\Lam, \Lam - \dl\Lam]$, where $\dl \Lam \ll \Lam$, as shown in \fig{fig:coarse-grain}.
The chemical potential remains unrenormalized, i.e. at perfect compensation, since the Hartree and Fock diagrams in \fig{fig:HF} vanish due to $\int_\Lam dK ~ \mc G(K) = 0$.
In the rest of this section we focus on the renormalizations to the  quartic vertices.

The combination of the UV cutoff imposed by $f_\Lam(k_r, k_z)$ in \eq{eq:action}, and conservation of momentum at the quartic vertices on the plane of the loop imposes strong kinematic constrains on  most scattering channels \cite{Shankar1994}.
Thus, instead of studying the complicated RG flow of entire coupling functions, we focus on the dominant scattering channels, which are identified by applying the kinematic constrains.
There are three scattering channels that dominate the low energy dynamics, 
\begin{itemize}
\item Pairing (BCS): $g_i(\{\what K_i\}) \rtarw  g_i(\what K_1, \what K_2, -\what K_1, -\what K_2) \equiv \Lam^{\eta - 2} V_i(\what K_1, \what K_2)$,
\item Small angle forward scattering (FS): $g_i(\{\what K_i\}) \rtarw  g_i(\what K_1, \what K_1, \what K_3, \what K_3) \equiv \Lam^{\eta - 2} U_i^{\fs}(\what K_1, \what K_3)$,
\item Large angle forward scattering (ES): $g_i(\{\what K_i\}) \rtarw  g_i(\what K_1, \what K_3, \what K_3, \what K_1) \equiv  \Lam^{\eta - 2} U_i^{\es}(\what K_1, \what K_3)$.
\end{itemize}
Since $g_i(\{\what K_i\})$ are dimensionful for $\eta\neq 2$, we have expressed the scaling dimension of $g_i(\{\what K_i\})$ in units of $\Lambda$, such that $V_i, U_i^{\fs},$ and $U_i^{\es}$ are dimensionless.
Since there are 4 types of interactions in \eq{eq:action}, the three channels generate $12$ coupling \textit{functions}. 
However, the FS and ES couplings are not truly distinct 
due to  non-conservation of pseudo-spin, and the interactions in the non-BCS, i.e. forward scattering,  channel can be represented either in terms of ES or FS couplings.
Here we adopt the FS representation, such that there are only eight {\it independent} coupling functions - four each for the BCS channel and the FS channel. 
As we will show below, the RG flow in the eight dimensional coupling space is further simplified by the fact that, at one-loop order, the flow of the BCS couplings are  decoupled from the flow of the FS couplings to the leading order in $\Lam/\kap \ll 1$. 
Additionally, owing to the $\theta$-rotation symmetry, the one-loop RG flow remains diagonal in the angular-momentum basis with identical flow for each harmonic.
This eliminates the complications arising from the functional nature of the couplings, since one may separately analyze the flow of coupling {\it constants} in a particular angular momentum channel, without worrying about coupling between different channels.

Because of its generic importance in the presence of extended zero-energy manifold in fermionic systems, we will first focus on the BCS channel, and then discuss the forward scattering channel in section \ref{sec:non-BCS} where exciton condensates arise.
In both sections \ref{sec:BCS} and \ref{sec:non-BCS} we derive the one-loop RG flow for the respective couplings, show their fixed point structure, and determine the trajectories of the RG flow towards strong-coupling.
We also identify the nature of the states that are realized at  strong-coupling by tuning a single parameter.
These states may be considered as finite coupling instabilities of the Weyl-loop semi-metal.


\section{RG analysis of BCS couplings} \label{sec:BCS}
\begin{figure}[!t]
\centering
\begin{subfigure}{0.45\columnwidth}
\includegraphics[width=0.95\columnwidth]{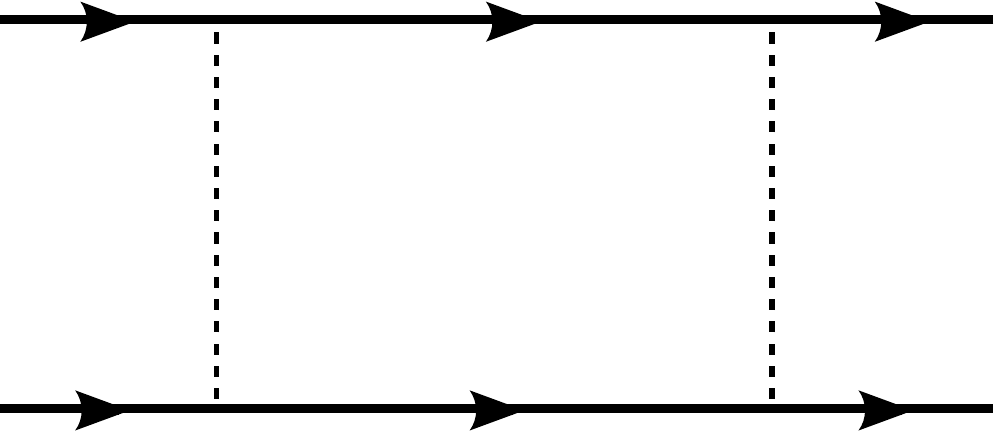}
\caption{$BCS$}
\label{fig:PP}
\end{subfigure}
\hfill
\begin{subfigure}{0.45\columnwidth}
\includegraphics[width=0.95\columnwidth]{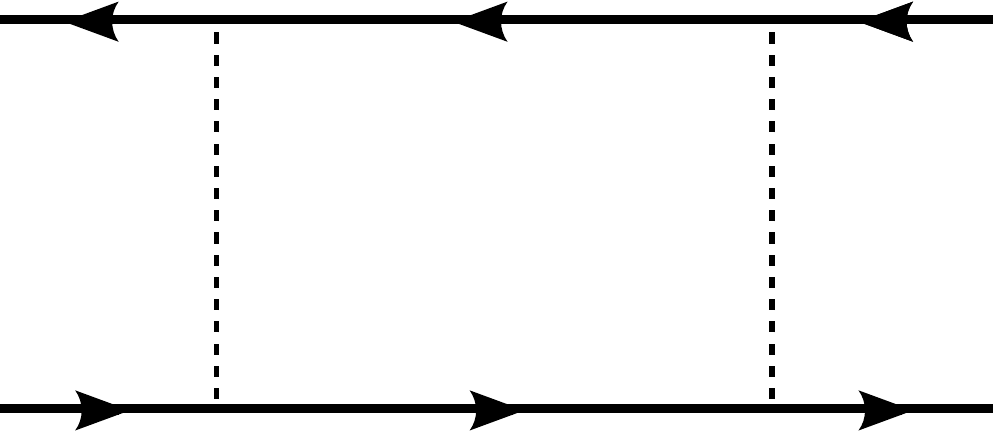}
\caption{$ZS'$}
\label{fig:PH}
\end{subfigure}
\hfill
\begin{subfigure}{0.45\columnwidth}
\includegraphics[width=\columnwidth]{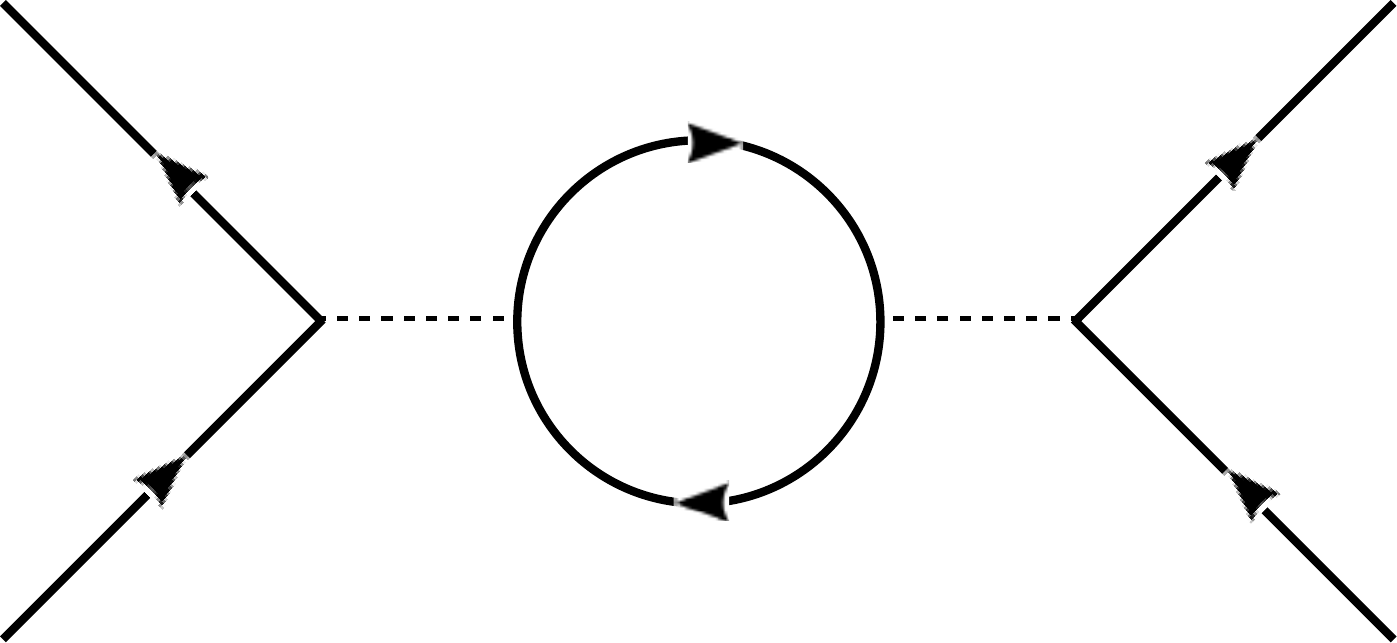}
\caption{$ZS$}
\label{fig:bubble}
\end{subfigure}
\hfill
\begin{subfigure}{0.45\columnwidth}
\includegraphics[width=0.95\columnwidth]{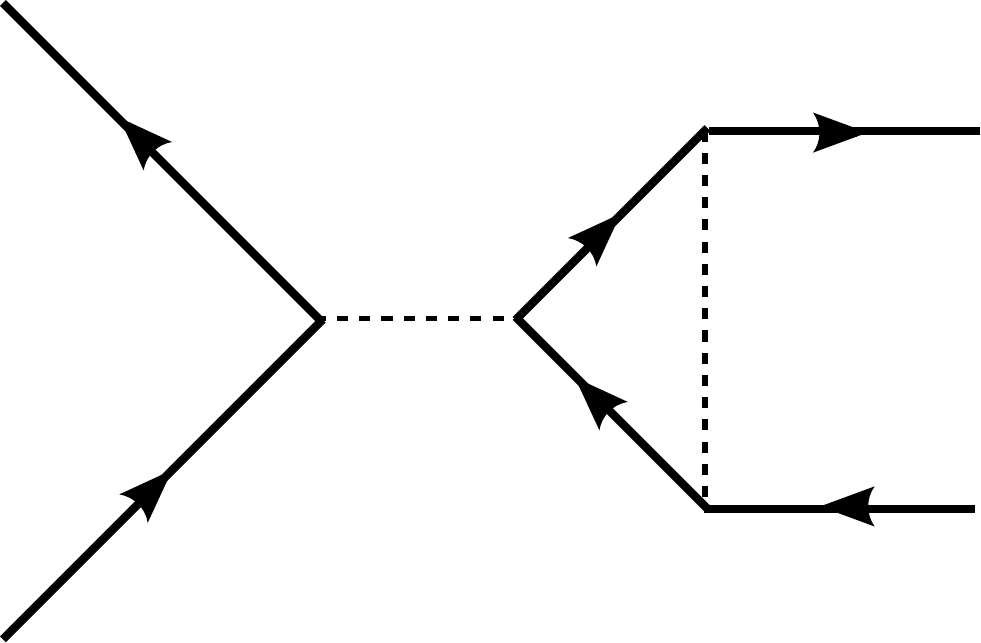}
\caption{$P$}
\label{fig:penguin}
\end{subfigure}
\caption{The four one-loop diagrams that renormalize the quartic vertex. We use the naming convention in \cite{Shankar1994} for (a), (b) and (c). Here, due to the matrix structure of the vertex, a fourth diagram is possible which we label as $P$ for ``penguin" diagrams.}
\label{fig:1L-diagrams}
\end{figure}

In this section we analyze the RG flow of the BCS couplings which are identified through the following kinematic constraint on the interaction vertices of the action,
\begin{widetext}
\begin{align}
S_{int}^{\bcs} &=  \Lam^{\eta-2}\int_\Lam \lt(\prod_{n=1}^4 dK_n \rt) ~(2\pi)^4 ~\dl^{(4)}(K_1 - K_2 + K_3 - K_4) ~ \dl(\what{\mbf{K}}_1 + \what{\mbf{K}}_3) ~\dl(\what{\mbf{K}}_2 + \what{\mbf{K}}_4) \nn \\
& \quad \times \Bigl[ \frac{V_{1}(\what K_1, \what K_2)}{\kap} \lt( \bar \psi(K_1) \gam_0  \psi(K_2) \rt) \lt( \bar \psi(K_3) \gam_0 \psi(K_4) \rt), \nn \\
& \qquad - \frac{V_{2+}(\what K_1, \what K_2)}{\kap} \Bigl\{\lt( \bar \psi(K_1) \gam_1 \psi(K_2) \rt) \lt( \bar \psi(K_3) \gam_1 \psi(K_4) \rt) + (\gam_1 \rtarw \gam_2) \Bigr\} \nn \\
& \qquad - \frac{V_{2-}(\what K_1, \what K_2)}{\kap} \Bigl\{\lt( \bar \psi(K_1) \gam_1 \psi(K_2) \rt) \lt( \bar \psi(K_3) \gam_1 \psi(K_4) \rt) - (\gam_1 \rtarw \gam_2) \Bigr\} \nn \\
& \qquad + \frac{V_{3}(\what K_1, \what K_2)}{\kap} \lt( \bar\psi(K_1)  \psi(K_2) \rt) \lt( \bar\psi(K_3)  \psi(K_4) \rt) \Bigr].
\label{eq:action-BCS}
\end{align}
\end{widetext}
Here $\what{\mbf{K}}_n$ is the unit vector along $\mbf{K}_n$.
Imposing the rotational invariance under the action of $\mc R_\theta$, constrains the functional form of $V_{j}(\what K_1, \what K_2) \mapsto V_{j}(\theta_1 - \theta_2)$, where $\theta_i$ are the angular positions on the loop and are  physically equivalent to $\what K_i$. 
We note that the two dimensional unit vector $\what K_i$ is the projection of the three dimensional unit vector $\what{\mbf{K}}_i$ on the plane of the loop; the dependence of $V_j$ on the third component of $\what{\mbf{K}}_i$ is irrelevant.
As we will show, different angular momentum channels further decouple, so that one may work with sets of coupling constants in a particular angular momentum channel.
Additionally, the action $S_{\eta,0} + S_{\bcs}$  with $V_{2+} = 0$ is  invariant under $\mc R_\vphi$ for  $\xi(\theta) =  \sgn{\theta}  \wtil{\xi}$ with $\wtil \xi \in (-\pi, \pi]$.
Since the symmetry involves a quasi-global choice of $\xi(\theta)$, it is a subgroup of the pseudospin-$U(1)_\infty$ symmetry, and we refer to it as BCS-$U(1)$ symmetry.
The BCS-$U(1)$ symmetry ensures that  $V_{2+}$ vertex is not generated by scatterings in the BCS channel, if it is absent at tree level.

There are four diagrams at one-loop order as shown in \fig{fig:1L-diagrams}.
The contribution from the $BCS$ diagram is enhanced by a factor of $\kap$ with respect to the other three diagrams.
This underscores the fact that the FS couplings do not mix with the BCS couplings at leading order in $\Lam/\kap$. 
We note further that the BCS diagram has a log divergence at $\epsilon = 0$ which makes the problem suitable for a RG analysis. 
While the analysis here is developed for the undoped material, this log divergence persists even at non-zero doping (whereupon it becomes just the familiar Cooper log). 
The situation in the doped case was discussed in \cite{Nandkishore2016} and is not discussed further here.
The RG flows of the BCS couplings at the leading order in $\Lam/\kap$ are given by
\begin{widetext}
\begin{align}
& \dow_\ell \V{1} = - \V{1} \lt[(2-\eta) + \frac{1}{4\pi} ( \V{1} + 2\V{2-} ) \rt] 
- \frac{1}{4\pi} ( \V{3}^2 + 2\V{2+}^2 + 2\V{2-}^2 + 2\V{3} \V{2-} ) 
\label{eq:flow-V1} \\
&\dow_\ell \V{2+} = - \V{2+} \Bigl[ (2-\eta) + \frac{1}{2\pi} (\V{1} - \V{3}) \Bigr],
\label{eq:flow-V2+} \\
&\dow_\ell \V{2-} 
= - (2-\eta) ~ \V{2-} - \frac{1}{8\pi} ~(2 \V{2-} + \V{1} + \V{3})^2,
\label{eq:flow-V2-} \\
& \dow_\ell \V{3} = - \V{3} \lt[(2-\eta) + \frac{1}{2\pi} ( \V{1} + \V{2-}) \rt] + \frac{1}{2\pi} ( \V{2+}^2 - \V{2-}^2 - \V{1} \V{2-} ). 
\label{eq:flow-V3} 
\end{align}
\end{widetext}
Here $\V{n}$ represents the $J$-th angular momentum harmonic of $V_n(\theta)$.
Since $\dow_\ell \V{2+} \propto -\V{2+}$, if bare $\V{2+} = 0$, then it is not generated during the course of the RG flow. 

\subsection{Fixed points} \label{sec:FP-SC}
\begin{figure}[!]
\centering
\begin{subfigure}{0.45\textwidth}
\includegraphics[width = 0.95\columnwidth]{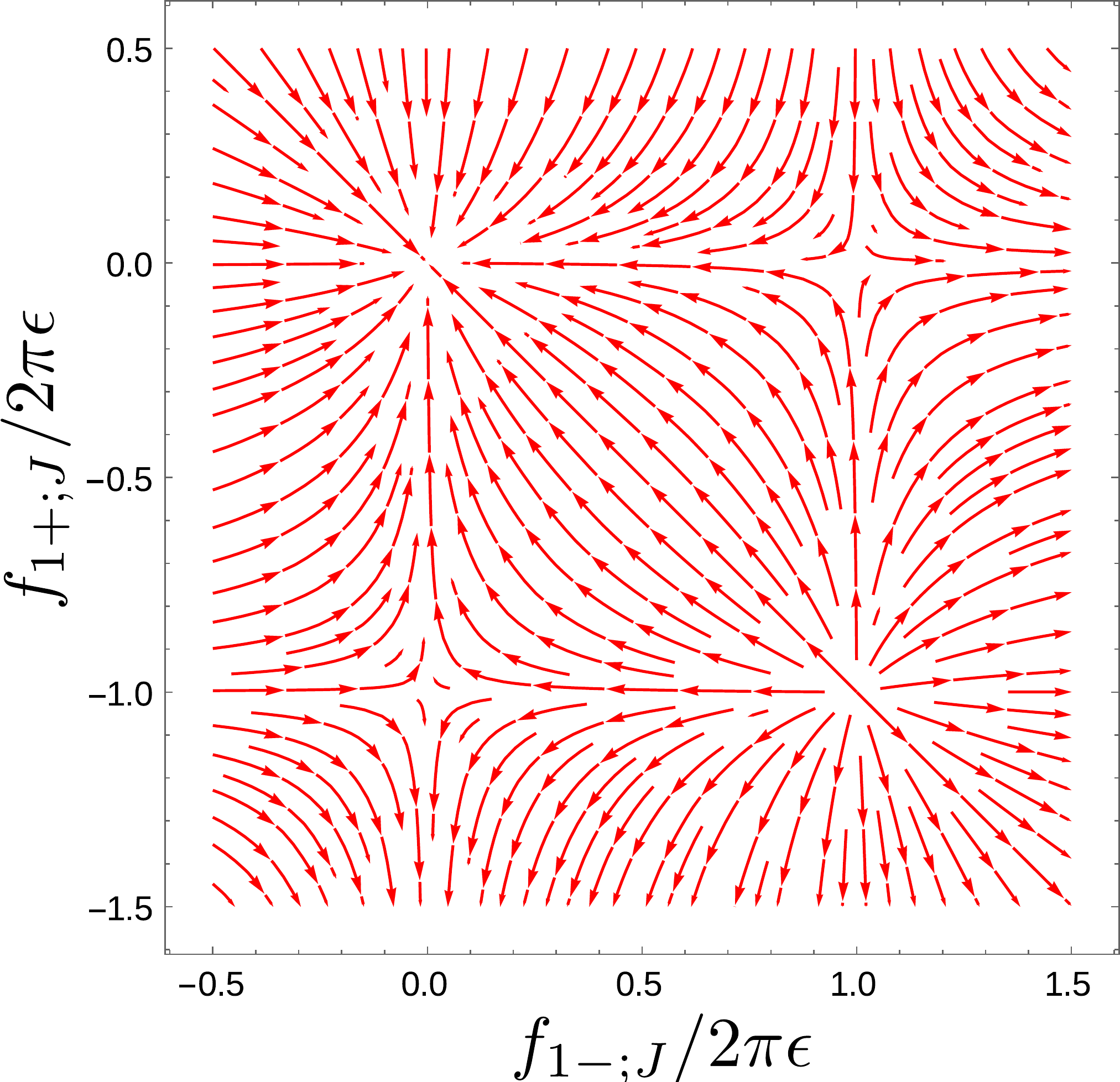}
\caption{}
\label{fig:flow-a}
\end{subfigure}
\hfill
\begin{subfigure}{0.45\textwidth}
\includegraphics[width = 0.95\columnwidth]{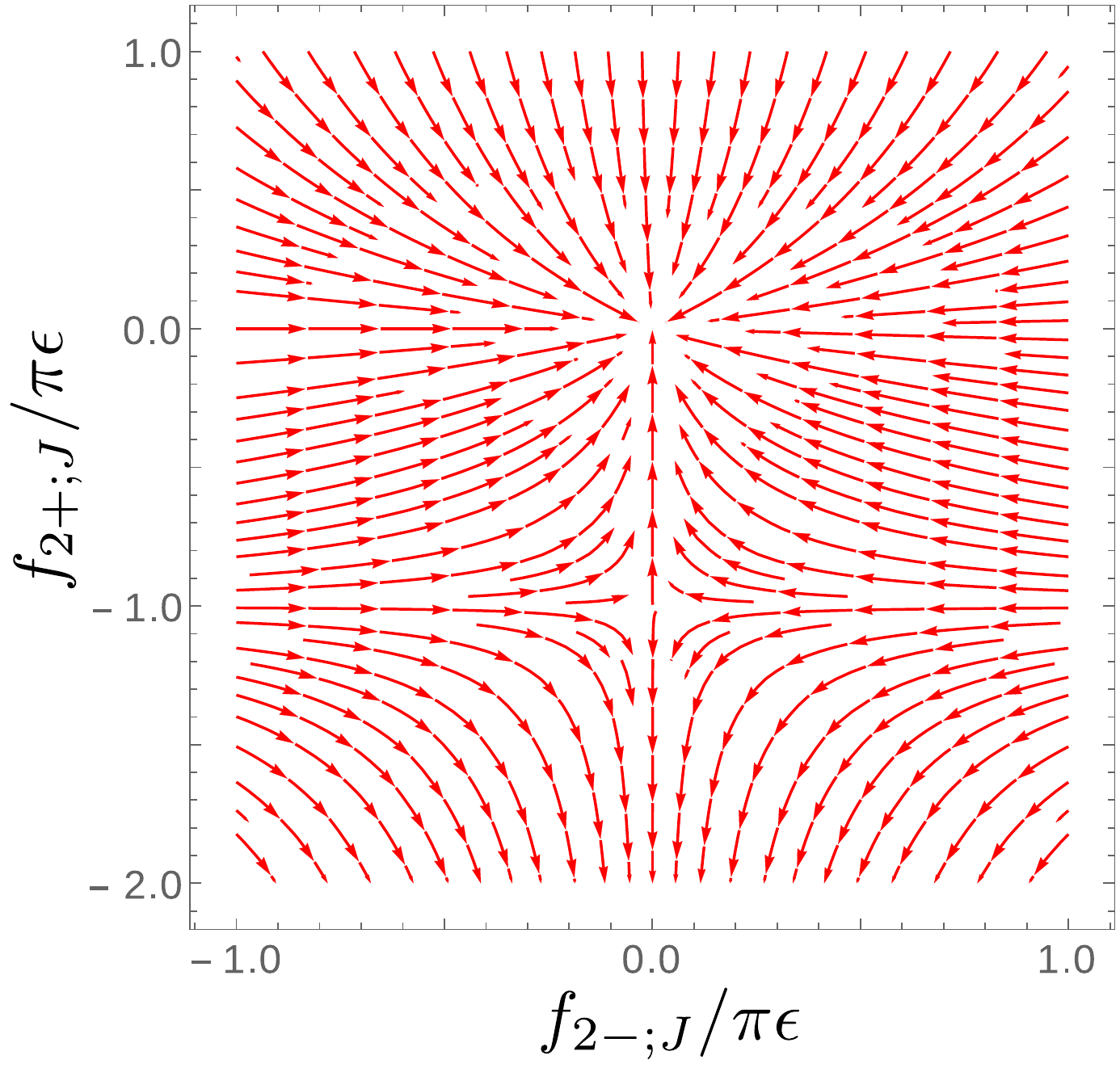}
\caption{}
\label{fig:flow-b}
\end{subfigure}
\caption{RG flow lines on (a) $(\f{1+}, \f{1-})$ and (b) $(\f{2+}, \f{2-})$ planes.}
\label{fig:flow}
\end{figure}
The expressions of the beta-functions are simplified by changing coordinates in the coupling space as 
\begin{align}
& \f{1\pm} = \V{2+} \pm \half (\V{1} - \V{3}), \nn \\
& \f{2\pm} = \V{2-} \pm \half (\V{1} + \V{3}).
\label{eq:fn}
\end{align}
The flows of $\f{n\pm}$ for $\eta = 2 - \eps < 2$ are governed by
\begin{align}
\dow_\ell \f{1+} 
&= - \f{1+} \Bigl[ \eps + \frac{1}{2\pi} \f{1+} \Bigr],
\label{eq:flow-f1+} \\
\dow_\ell \f{1-} 
&= - \f{1-} \Bigl[ \eps - \frac{1}{2\pi} \f{1-} \Bigr],
\label{eq:flow-f1-} \\
\dow_\ell \f{2+} 
&= - \f{2+} \Bigl[ \eps + \frac{1}{\pi} \f{2+} \Bigr],
\label{eq:flow-f2+} \\
\dow_\ell \f{2-} 
&= - \eps \f{2-}.
\label{eq:flow-f2-}
\end{align}
The beta-functions imply that $\f{n\pm}$ do not mix at one-loop order.
We plot the projections of the four dimensional RG flow on the $(\f{1+}, \f{1-})$ and $(\f{2+}, \f{2-})$ planes in \fig{fig:flow}.
The fixed points are derived from the conditions for simultaneous vanishing of the four beta functions, which result in 3 quadratic and 1 linear equations that have $2^3$ solutions.
The Gaussian fixed point is the only stable fixed point of the RG flow.
It has a finite basin of attraction, whose volume is controlled by $\eps$.
The non-Gaussian fixed points have at least one relevant direction which take(s) the flow towards the Gaussian fixed point or strong-coupling, depending on which side of the sepatrices the couplings lie.
In Table \ref{tab:FP-BCS-1}  we list all non-Gaussian fixed points  according to the number of relevant directions they possess.
\begin{table}[!t]
\centering
\begin{tabular}{c || c | c | c | c || c}
\# & $\f{1+}^*$ & $\f{1-}^*$ & $\f{2+}^*$ & $\f{2-}^*$ & Tune \\
\hline \hline 
\rom{1}$_{\bcs}$ & $-2$ & $0$ & $0$ & $0$ 
& $f_{1+}$ \\
\hline 
\rom{2}$_{\bcs}$ & $0$ & $2$ & $0$ & $0$ 
& $f_{1-}$ \\
\hline 
\rom{3}$_{\bcs}$ & $0$ & $0$ & $-1$ & $0$ 
& $f_{2+}$ \\
\hline 
\rom{4}$_{\bcs}$ & $-2$ & $2$ & $0$ & $0$ 
& $f_{1\pm}$ \\
\hline 
\rom{5}$_{\bcs}$ & $-2$ & $0$ &$-1$ & $0$ 
& $f_{1+}, f_{2+}$ \\
\hline 
\rom{6}$_{\bcs}$ & $0$ & $2$ & $-1$ & $0$
& $f_{1-}, f_{2+}$ \\
\hline 
\rom{7}$_{\bcs}$ & $-2$ & $2$ & $-1$ & $0$
& $f_{1\pm}, f_{2+}$ \\
\hline
\end{tabular}
\caption{Non-Gaussian fixed points in units of $\pi \eps$ for the BCS couplings, with at least one relevant direction. 
The number of relevant couplings required to be tuned to achieve the criticality determines its order, viz. stable (0), critical (1), bicritical (2), and tricritical (3), where the numbers within the parentheses are the number of relevant couplings.
Thus, fixed points \rom{1}$_\bcs$ - \rom{3}$_\bcs$ are critical,  \rom{4}$_\bcs$ - \rom{6}$_\bcs$ are bicritical, and \rom{7}$_\bcs$ is tricritical.}
\label{tab:FP-BCS-1}
\end{table}

In order to interpret the fixed points  in terms of the original couplings of the model we invert the relation \eq{eq:fn} to obtain
\begin{align}
& \V{1} =  \half ((\f{2+} - \f{2-}) + (\f{1+} - \f{1-}) ), 
\nn \\
& \V{2+} = \half ( \f{1+} + \f{1-}), 
\nn \\
& \V{2-} = \half ( \f{2+} + \f{2-}), \nn \\
& \V{3} =  \half ((\f{2+} - \f{2-}) - (\f{1+} - \f{1-}) ).
\label{eq:old-couplings}
\end{align}
From Table \ref{tab:FP-BCS-1} we note that at \rom{3}$_\bcs$ $\V{2+}$ vanishes, which implies the emergence of the BCS-$U(1)$ symmetry at the critical fixed point.
The BCS-$U(1)$ symmetry is also present at the bi- and tri-critical fixed points \rom{4}$_\bcs$ and \rom{7}$_\bcs$, respectively.
For the rest of this section we focus on the subspace where $\V{2+} = 0$, i.e. the subspace invariant under BCS-$U(1)$. To motivate this approximation, note that it is natural to take the bare UV scale interaction to be pure density-density, without any pseudospin structure. An interacting theory with only density-density interactions will have this BCS-$U(1)$ symmetry. Interactions with non-trivial pseudospin structure will then be generated under the RG, but the only those interactions that lie within the `maximally symmetric subspace.' In particular, $V_{2+}$, which breaks the BCS-$U(1)$ symmetry, will not be generated. Additionally of course, restricting to the maximally symmetric subspace has the advantage of providing us with a `toy model' that is more amenable to analysis.  
\begin{table}[!t]
\centering
\begin{tabular}{c || c | c | c | c }
\#  & $\V{1}^*$ & $\V{2-}^*$ & $\V{3}^*$ & Tune\\
\hline \hline 
\rom{3}$_{\bcs}$  & $-1/2$  & $-1/2$ & $-1/2$ & $f_{2+}$\\
\hline 
\rom{4}$_{\bcs}$  & $-2$  & $0$ & $2$ & $(V_1-V_3)$ \\
\hline 
\rom{7}$_{\bcs}$  & $ - 5/2$ & $-1/2$ & $3/2$ & $(V_1-V_3)$, $f_{2+}$\\
\hline
\end{tabular}
\caption{
The fixed points (in units of $\pi \eps$) in Table \ref{tab:FP-BCS-1} that lie in the $\V{2+}=0$ subspace.
The first two are critical fixed points, while the last one is bicritical.
}
\label{tab:FP-BCS-2}
\end{table}

We therefore restrict ourselves to the subspace with $V_{2+} -0$. Since $\V{2+}$ was a relevant perturbation at \rom{4}$_{\bcs}$ and \rom{7}$_{\bcs}$, these fixed points  become critical and bicritical, respectively, with respect to BCS-$U(1)$ invariant perturbations.
Thus, we obtain one Gaussian, two critical, and one bi-critical fixed points.
The non-Gaussian fixed points are  listed in Table \ref{tab:FP-BCS-2} in terms of  $\V{n}$.

Since both \rom{3}$_\bcs$ and  \rom{4}$_\bcs$ are critical fixed points, they potentially separate the non-interacting Gaussian fixed point (Weyl-loop semi-metal phase) from superconducting phases. 
We first discuss the stability of \rom{3}$_\bcs$, and then apply the same analysis to \rom{4}$_\bcs$. 
Since the RG flow of $\f{n\pm}$ mutually decouple and they are irrelevant when $|\f{n\pm}| \ll \eps$, the critical fixed point \rom{3}$_\bcs$, which is realized at $\f{1\pm}, \f{2-} = 0$, can only be destabilized by perturbations with  non-zero component along  $\f{2+}$.
As depicted in \fig{fig:flow-b},  the sign of the deviation $\dl \f{2+} = \f{2+} - \conj{\f{2+}}$ determines whether the perturbation takes the flow towards the Gaussian fixed point or  towards strong coupling where $\f{2+}$ is large and \textit{negative}.

In contrast to \rom{3}$_\bcs$, \rom{4}$_\bcs$ is located on the $(\f{1+}, \f{1-})$ plane with $\f{2\pm} = 0$. 
Since in the $\V{2+} = 0$ subspace $\f{1\pm}$ are equivalent, the RG flow in the neighborhood of \rom{4}$_\bcs$ is governed by
\begin{align}
& \dow_{\ell} \dl \f{1-} = \eps \dl \f{1-}, \quad
\dow_{\ell} \dl \f{2\pm} = - \eps \dl \f{2\pm}.
\end{align}
Therefore, perturbations with $\dl \f{1-} \neq 0$ are relevant and, depending on its sign, take the flow either towards the Gaussian fixed point, or towards strong coupling where $\f{1-}$ is large and \textit{positive}. 

\subsection{Flow to strong coupling} \label{sec:SCT}
%
In this subsection we identify the effective interactions along the stable RG flow trajectory that takes the theory towards strong coupling, as we tune away from the critical fixed points.
As in the preceding subsection, we discuss the flow away from \rom{3}$_\bcs$ first, followed by \rom{4}$_\bcs$. 

Let us label the flow away from the critical point towards strong coupling on the $\f{2+}$ axis as a  strong coupling trajectory (SCT).
Due to the stability of the Gaussian fixed point in the $(\f{1+}, \f{1-}, \f{2-})$ subspace, the SCT is stable against pertubations perpendicular to it.
With the aid of \eq{eq:old-couplings} we note that on the SCT $\V{1} = \V{2-} = \V{3} = \half \f{2+} < -\frac{\pi \eps}{2}$.
Thus, the BCS vertices on the SCT are given by
\begin{widetext}
\begin{align}
L_{\rom{3}}^{\bcs}(\{K_n\}) &=  \frac{1}{\kap} V_{1}(\hat K_1, \hat K_2) \lt[  (\bar \psi \gam_0 \psi)^2 + (\bar \psi \psi)^2 -  (\bar \psi \gam_1 \psi)^2 + (\bar \psi \gam_2 \psi)^2 
\rt],
\end{align}
where we have suppressed the momentum dependence of the fields.
On the plane of the Weyl-loop the vertices simplify to
\begin{align}
L_{\rom{3}}^{\bcs}(\vec K_1, \vec K_2) &=  \frac{1}{2\kap} V_{1}(\hat K_1, \hat K_2) 
~ \lt\{ \psi^{\dagger}(\vec K_1) (\psi^{\dagger}(-\vec K_1))^{\intercal} \rt\}
\lt\{ \psi^{\intercal}(-\vec K_2) \psi(\vec K_2) \rt\}.
\label{eq:SCT-BCS-1}
\end{align}
Since $V_{1} < 0$ on the SCT, this indicates that a pairing instability is driven by the condensation of  $\psi^{\intercal}(-\vec K) \psi(\vec K)$.

The SCT originating at \rom{4}$_\bcs$ is defined by $(\V{1}, \V{2\pm}, \V{3}) = (-\f{1-}, 0, \f{1-})$. 
Therefore, along the SCT the BCS vertices with momenta on the plane of the Weyl-loop simplifies to 
\begin{align}
L_{\rom{4}}^{\bcs}(\vec K_1, \vec K_2) &= V_1(\hat K_1, \hat K_2) 
\lt[ 
\lt\{ \psi^{\dagger}(\vec K_1) ~ \sig_1 ~ (\psi^{\dagger}(-\vec K_1))^{\intercal} \rt\}  
\lt\{ \psi^{\intercal}(-\vec K_2) \sig_1 \psi(\vec K_2) \rt\} \rt. \nn \\
&  \qquad \qquad \lt.
+ \lt\{ \psi^{\dagger}(\vec K_1) ~\sig_2 ~ (\psi^{\dagger}(-\vec K_1))^{\intercal} \rt\}
\lt\{ \psi^{\intercal}(-\vec K_2) ~\sig_2 ~ \psi(\vec K_2) \rt\} 
\rt].
\label{eq:SCT-BCS-2}
\end{align}
\end{widetext}
Since $V_1 < 0$ along the SCT, both vertices in \eq{eq:SCT-BCS-2} can drive a pairing instability.
In the following subsection we   verify the identity of the superconducting states indicated above through explicit computation of the anomalous scaling dimension of various pairing  susceptibilities along the two SCTs.


\subsection{Symmetry broken states} \label{sec:suscep}
In this subsection we determine the nature of the superconducting states that arise as instabilities of the critical points in the $\V{2+} =0$ subspace.
In particular, we compute the change of scaling dimension of the pairing susceptibilities along the SCT  as the system flows towards strong-coupling.

\begin{figure}[!t]
\centering
\includegraphics[scale=0.7]{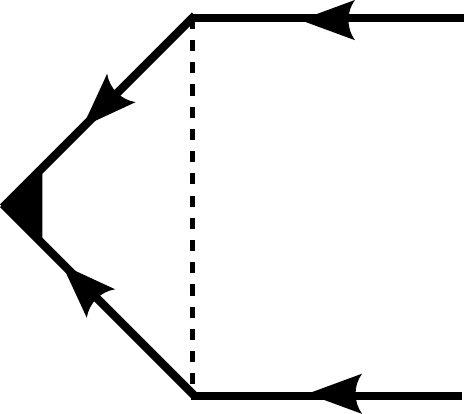}
\caption{}
\label{fig:suscep}
\end{figure}
We consider insertions of fermion pairs,
\begin{align}
S_2^{(\mu)} = \int dK ~ \conj{\Dl_\mu}(\what K) ~ \Psi^{\intercal}(-K) ~\gam_\mu ~\Psi(K) + \mbox{h.c.},
\label{eq:insert}
\end{align}
where $\mu = 0, 1, 2, 3$.
The pairing amplitude $\Dl_\mu(\what K)  \gam_\mu = - \Dl_\mu(-\what K) \gam_\mu^{\intercal}$, with $\Dl_\mu(\what K)$ being a complex valued function.
The `singlet' pairing corresponds to $\gam_\mu^{\intercal} = -\gam_\mu$ and  $\Dl_\mu(\hat K) = \Dl_\mu(-\hat K)$, while for $\gam_\mu^{\intercal} = \gam_\mu$ and $\Dl_\mu(\hat K) = - \Dl_\mu(-\hat K)$ the pairing occurs in a `triplet' channel. 
At one-loop order  $\Dl_\mu(\what K)$ is renormalized by \fig{fig:suscep}.
As derived in appendix \ref{app:chi} the RG flow of each angular momentum harmonic of  $\Dl_\mu(\what K)$ is governed by
\begin{align}
\dow_\ell \D{\mu} = \D{\mu} \lt[ 2 + \dl_{\mu}(\{\V{i}\}) \rt],
\label{eq:flow-Dl}
\end{align} 
where the anomalous dimension of $\D{\mu}$, $\dl_{\mu}(\{\V{i}\})$, is defined in \eq{eq:dl-BCS}. 

Along the SCT originating from  \rom{3}$_\bcs$ the susceptibility for $\gam_3$ is most strongly enhanced, and the RG flow of $\Dl_3(\what K)$ is governed by
\begin{align}
\dow_\ell \D{3} = \D{3}
\lt[ 2 - \frac{\V{1}}{\pi} \rt],
\end{align}
where we have used the fact that along the SCT $\V{1} = \V{3} = \V{2-} = \half \f{2+}$.
Since $\V{1} < 0$, $\Dl_3(\what K)$ obtains a positive anomalous dimension.
The scaling dimensions of $\Dl_\mu(\what K)$ for $\mu \neq 3$ do not change because  $\dl_{\mu \neq 3} = 0$.

From the symmetry properties of the gap function, we obtain $\Dl_3(-\what K)= - \Dl_3(\what K)$, which is equivalent to $\Dl_3(\theta+\pi)= - \Dl_3(\theta)$ in terms of the loop  coordinate $\theta$.
Decomposing $\Dl_3(\theta)$ in terms of angular momentum harmonics,
\begin{align}
\Dl_3(\theta) = \sum_{J=-\infty}^{\infty} e^{-i \theta J} ~ \D{3},
\end{align}
we note that only the odd $J$ harmonics are non-zero. While the flow equations in different odd angular momentum channels are identical, the degeneracy between different angular momentum channels will be broken by the initial conditions. The largest `bare interaction' is likely to arise in the lowest allowed angular momentum channel (i.e. $J = \pm1$) which will then be the leading instability. 
Therefore, the leading instability is expected to be `p-wave' consistent with general arguments \cite{Nandkishore2016}. 
Retaining only the $J=\pm 1$ harmonic we express
\begin{align}
\Dl_3(\theta) &=  \wtil\Dl_+ \cos{\theta} 
+ i \wtil\Dl_- \sin{\theta},
\label{eq:Dl3}
\end{align}
where the constants $\wtil \Dl_{\pm} = 2(\Dl_{3;-1} \pm \Dl_{3;1})$.
Allowing for weak radial momentum dependence on the plane of the loop we generalize $\Dl_3(\theta)$ to
\begin{align}
\Dl_3(\vec K) =  \wtil \Dl_+ ~ \frac{K_x}{\kap} + i \wtil \Dl_- ~ \frac{K_y}{\kap}.
\end{align}
In terms of the generalized expression for a superconducting order parameter 
\begin{align}
\tensor{\Dl}(\vec K) = i (d_0(\vec K) + \mbf d(\vec K) \cdot \mbf \sigma) \sig_2,
\label{eq:general-Dl}
\end{align}
the current state corresponds to $d_0(\vec K), d_1(\vec K), d_3(\vec K) = 0$, and $d_2(\vec K) = -i \Dl_3(\vec K)$.

Note that we have two degenerate channels $J=\pm 1$ which are related by time reversal symmetry. We now discuss the competition of these two channels below $T_c$. 
For the superconducting state where $J=\pm 1$ components are  `in-phase' ($\Dl_{3;1} = \Dl_{3;-1}$) and `out-of-phase' ($\Dl_{3;1} = - \Dl_{3;-1}$),  $(\wtil\Dl_+, \wtil\Dl_-)  = (4 \Dl_{3;-1}, 0)$ and $(\wtil\Dl_+, \wtil\Dl_-)  = (0,4 \Dl_{3;-1})$, respectively.
Therefore, in this state the gap function vanishes at two points on the Fermi surface - this is nodal superconductivity which spontaneously breaks rotation symmetry. In contrast if only one out of the $J_{\pm}$ channels develops a non-zero order parameter this corresponds to 
\begin{align}
\Dl_3(\vec K) =  
\begin{cases}
\dfrac{2 \Dl_{3;-1}}{\kap} ~(K_x + i K_y) & \mbox{ for } J = -1 \\[4ex]
- \dfrac{2 \Dl_{3;1}}{\kap} ~(K_x - i K_y) & \mbox{ for } J = +1
\end{cases}
\label{eq:p-wave}
\end{align}
This type of ordering spontaneously breaks time reversal symmetry and corresponds to chiral superconductivity. Note that these gap functions do not vanish on the Fermi surface, and thus are expected to have a larger condensation energy. 
We show this explicitly by minimizing the Ginzburg-Landau free energy, similar to \cite{graphene1, graphene2, graphene3, graphene4, graphene5}. 

Since $V_1 < 0$ along the SCT, we introduce an auxillary field, $\phi(k)$, to decompose the first term in \eq{eq:SCT-BCS-1}. 
Integrating out the fermions generates an effective Ginzburg-Landau action for $\phi(k)$.
In the symmetry broken state below the critical temperature we ignore spatiotemporal fluctuations of $\phi(k)$, and focus on the `potential' part of the effective action.
We ignore contributions from scattering between Cooper pairs at different parts of the loop, and express the total free energy as a sum over free energy per unit length of the loop,
\begin{align}
\mc F = \int_{-\pi}^{\pi} \frac{d\theta}{2\pi}~ \mc{F}'(\theta) =  \int_{-\pi}^{\pi} \frac{d\theta}{2\pi}~ \lt( a|\phi(\theta)|^2 + b |\phi(\theta)|^4 \rt),
\end{align}
where $a$ and $b$ are effective parameters.
Substituting $\phi(\theta) = \Dl_3(\theta)$ (defined in \eq{eq:Dl3}) leads to
\begin{widetext}
\begin{align}
\mc F = \frac{a}{2} \lt( |\wtil \Dl_+|^2 + |\wtil \Dl_-|^2\rt)
+ \frac{3b}{8} \lt(|\wtil \Dl_+|^4 + |\wtil \Dl_-|^4 \rt) 
+ \frac{b}{4} |\wtil \Dl_+|^2 |\wtil \Dl_-|^2 \lt( 1 - \frac{1}{2} \lt( \frac{\conj{\wtil\Dl_+} \wtil\Dl_-}{|\wtil \Dl_+| |\wtil \Dl_-|} - \mbox{c.c.}  \rt)^2 \rt).
\label{eq:totF}
\end{align}
\end{widetext}
We note that the third term represents a repulsion between the two components of $\Dl_3(\theta)$.
For a condensate to form the condensation energy needs to overcome the energy barrier due to the repulsion.
Since in the superconducting phase $a < 0$ and $b>0$, the configuration that minimizes $\mc F$ is determined by the relative magnitude of $|a|$ and $b$.
By expressing $(\wtil \Dl_+,\wtil \Dl_-) = \wtil \Dl_+ (1, \sqrt{x} e^{i A} )$, in units of $b |\wtil \Dl_+|^4$ (or equivalently $b |\wtil \Dl_-|^4$) we obtain  
\begin{align}
\frac{2\mc F}{b |\wtil \Dl_+|^4} 
&= - \frac{|a|}{b |\wtil \Dl_+|^2} ( 1 + x) 
+ \frac{3}{4} \lt(1 - x \rt)^2 \nn \\
& \qquad + 2 x \lt( 1 + \frac{1}{2} \sin^2{A} \rt).
\end{align}
Minimization of the scaled free energy with respect to the relative magnitude $x$ and the relative phase $A$ leads to three possible states corresponding to $x=0$, $(x > 0, A = \pi/2 \mbox{ or }  3\pi/2)$, and $(x > 0, A = 0 \mbox{ or }  \pi)$ as long as $\frac{|a|}{b |\wtil \Dl_+|^2} > 1/2$.
The conditional inequality is selfconsistently satisfied by  $|\wtil \Dl_+|$ at each minimum.
The first two minima correspond to nodal states, while the last one is a pair of nodeless states which are  equivalent to those in \eq{eq:p-wave} for $x=1$.
The lower bound on the dimensionless ratio $\frac{|a|}{b |\Dl_+|^2}$ originates from the competition between the repulsion  and condensation energy.
By comparing the free energy at these minima we conclude that the nodeless state is realized at  the  global minimum of the free energy.
Thus the leading instability associated with the flow to strong coupling emerging from the \rom{3}$_\bcs$ vertex is to a fully gapped chiral state with odd angular momentum and with $\Delta(\vec{k})$ proportional to the {\it unit matrix} in pseudospin space (i.e. $d_1=d_3=0$). 
Further, we note that this is the only state that involves solely  intra-band pairing, and is smoothly connected to paired states in both the conduction and valence band.
Thus this state is expected to be  most robust to chemical potential disorder \cite{Nandkishore2016}. Indeed, it may even be {\it enhanced} by disorder through the mechanism discussed in \cite{enhance1, enhance2, enhance3}.

Applying the above analysis to the SCT originating from \rom{4}$_\bcs$, we find that pairing susceptibilities for $\gam_1$ and $\gam_2$ vertices are enhanced identically, while $\gam_0$ and $\gam_3$ are unaffected. 
From the symmetry of $\Dl_\mu \gam_\mu$ we identify the $\gam_1$ ($\gam_2$) pairing vertex as a singlet (triplet).
The triplet pairing is distinguished from the one associated with \rom{3}$_\bcs$ with the aid of \eq{eq:general-Dl}, and it corresponds to $d_{\mu \neq 3}(\what K) = 0$ and $d_{3}(\what K) = \Dl_3(\what K)$.
The singlet corresponds to $\mbf d(\what K) = 0$ and $d_0(\what K) = - i \Dl_1(\what K)$. 
While the quantum scaling dimensions of the singlet and triplet pairings are identical  due to the BCS-$U(1)$ symmetry, the pairings occur in distinct angular momentum channels: 
the singlet (triplet) pairing occurs in even (odd) angular momentum channel. 
Since there is no reason why the bare couplings (which set the initial conditions for the RG flow) should be equal in distinct angular momentum channels, the apparent degeneracy will thus be broken by the initial conditions, 
 and the leading instability will occur in the channel $d_{\mu \neq 3}(\what K) = 0$ and $d_{3}(\what K) = \Dl_3(\what K)$ if the most attractive bare coupling is an odd angular momentum channel, and in the channel $\mbf d(\what K) = 0$ and $d_0(\what K) = - i \Dl_1(\what K)$ if the most attractive bare coupling is in an even angular momentum channel. 
In the case where the leading instability is in a channel with non-zero $J$, the $\pm J $ channels will again be degenerate, and one may have either fully gapped chiral superconductors or gapless non-chiral superconductors. 
An analysis of the most likely symmetry broken state resulting from the instability driven by the $\gam_2$ pairing vertex indicates a fully gapped $p$-wave state as obtained above.
However, it is distinguished from the same through the nontrivial matrix structure of the order parameter in the pseudospin space since $d_3 \neq 0$.
The most likely candidate for the symmetry broken state for the singlet pairing is a uniform s-wave superconductor.
We note that these states involve {\it interband} pairing \cite{Nandkishore2016} and thus will likely be rapidly disrupted by chemical potential disorder, unlike the state arising from the flow out of \rom{3}$_\bcs$. 


\section{RG analysis of forward scattering channel} \label{sec:non-BCS}
In this section we discuss the RG flow of the forward scattering channel.
In the absence of nesting, condensation of intra-orbital  particle-hole pairs carrying a finite momentum is suppressed by a lack of density of states.
Consequently, additional fine tuning is necessary to drive such a phase transition.
Thus in a single-orbital system the forward scattering channel does not lead to a weak coupling  instability of the metallic state \cite{Shankar1994}.
However, in multi-orbital systems  additional forward scatterings between different orbitals are present, which can lead to the condensation of inter-orbital particle-hole pairs which carry zero net momentum.
In the presence of a Fermi surface or nodal lines, the zero-momentum pairing of electrons and holes can utilize the extended manifold of degenerate states available at the Fermi level to enhance their condensation energy.
Another way to say this is to note that there is a log divergence in the forward scattering channel for the undoped Weyl loop system (at $\eta = 2$) which can lead to an excitonic instability. 
Since the exciton condensation crucially depends on the degenracy of the two bands, this log divergence is cut off by doping  and there is no weak coupling instability for the torus Fermi surface. 
However our focus here is on the possible symmetry broken phases resulting from instabilities driven by  forward scatterings in the undoped system. 

As noted earlier, there are two equivalent ways of representing the interaction vertices for the forward scattering channel.
Here we have adopted the FS representation, and express the vertices as
\begin{widetext}
\begin{align}
S_{int}^{\fs} &= \Lam^{\eta-2} \int_\Lam \lt(\prod_{n=1}^4 dK_n \rt) ~(2\pi)^4 ~\dl^{(4)}(K_1 - K_2 + K_3 - K_4) ~ \dl(\what{\mbf{K}}_1 - \what{\mbf{K}}_2) ~\dl(\what{\mbf{K}}_3 - \what{\mbf{K}}_4) \nn \\
& \quad \times \Bigl[ \frac{U_{1}(\what K_1, \what K_3)}{\kap} \lt( \bar \psi(K_1) \gam_0  \psi(K_2) \rt) \lt( \bar \psi(K_3) \gam_0 \psi(K_4) \rt), \nn \\
& \qquad - \frac{U_{2+}(\what K_1, \what K_3)}{\kap} \Bigl\{\lt( \bar \psi(K_1) \gam_1 \psi(K_2) \rt) \lt( \bar \psi(K_3) \gam_1 \psi(K_4) \rt) + (\gam_1 \rtarw \gam_2) \Bigr\} \nn \\
& \qquad - \frac{U_{2-}(\what K_1, \what K_3)}{\kap} \Bigl\{\lt( \bar \psi(K_1) \gam_1 \psi(K_2) \rt) \lt( \bar \psi(K_3) \gam_1 \psi(K_4) \rt) - (\gam_1 \rtarw \gam_2) \Bigr\} \nn \\
& \qquad + \frac{U_{3}(\what K_1, \what K_3)}{\kap} \lt( \bar\psi(K_1)  \psi(K_2) \rt) \lt( \bar\psi(K_3)  \psi(K_4) \rt) \Bigr],
\label{eq:action-ES}
\end{align}
\end{widetext}
where $U_{j}(\what K_1, \what K_3) \equiv U_{j}(\theta_1 - \theta_3)$, and we have dropped explicit reference to the representation for the coupling functions.
Due to the kinematic restriction, the charge-$U(1)_\infty$ symmetry is present in the forward scattering sector.
Indeed $S_{0;\eta} + S_{int}^{\fs}$ resembles the Fermi liquid fixed point. 
Although $\mc R_\vphi$ is not a symmetry of $S_{int}^\fs$ for any non-trivial choice of $\xi(\theta)$, it becomes a symmetry when $U_{2-} = 0$ with $\xi(\theta) = \wtil \xi$.
In order to contrast a similar symmetry present in the BCS sector, we refer to the current one as the FS-$U(1)$ symmetry.
The RG flow in the $U_{2-} = 0$ subspace is protected by the FS-$U(1)$ symmetry, which implies that $\dow_\ell U_{2-} \propto U_{2-}$.

\subsection{Fixed points}
In the forward scattering channel, even at zero energy, the net incoming momentum is generically non-zero as the momenta of typical incoming states are not anti-parallel in $\mbf K$-space.
In order to transfer the finite  momentum of the incoming states to the outgoing states, the  virtual excitations must carry a net  momentum.
Therefore, scattering processes that favor virtual exciations with zero net momentum  are suppressed for forward scattering channels at low energy, as is the case for the $BCS$ diagram.
The internal loop in the other three diagrams in \fig{fig:1L-diagrams} carry a net momentum, and renormalizes $U_{j}$ at leading order in $\Lam/\kap$.
The FS couplings flow according to
\begin{align}
& \dow_{\ell} \U{1} = - \eps \U{1}, 
\label{eq:flow-FS-1}
\\
& \dow_{\ell} \U{2+} = 
- \U{2+} \lt[ \eps + \frac{2}{\pi} \U{2+} \rt] -  \frac{2}{\pi} \U{2-}^2,
\label{eq:flow-FS-2+}
\\
& \dow_{\ell} \U{2-} = 
- \U{2-} \lt[ \eps + \frac{4}{\pi}  \U{2+} \rt], 
\label{eq:flow-FS-2-} \\
& \dow_{\ell} \U{3} = - \U{3} \Bigl[ \eps + \frac{4}{\pi} \U{3} \Bigr].
\label{eq:flow-FS-3}
\end{align}
It is interesting to note that when all four couplings are repulsive, they are irrelevant. 
Moreover, the $U_1$ vertex which mediates scatterings between total densities in momentum space, $\sum_{i=1,2} c_i^{\dag}(\mbf K)  c_i(\mbf K)$, remains unrenormalized.


\begin{table}[!t]
\centering
\begin{tabular}{c || c | c | c | c || c}
\# & $\U{1}^*$ & $\U{2+}^*$ & $\U{2-}^*$ & $\U{3}^*$ &  Tune \\
\hline \hline 
\rom{1}$_{\fs}$ & $0$ & $0$ & $0$ & $-1/4$ 
& $\U{3}$ \\
\hline 
\rom{2}$_{\fs}$ & $0$ & $-1/4$ & $-1/4$ & $0$  
& $(\U{2+}+\U{2-})$ \\
\hline 
\rom{3}$_{\fs}$ & $0$ & $-1/4$ & $1/4$ & $0$ 
& $(\U{2+}-\U{2-})$ \\
\hline 
\rom{4}$_{\fs}$ & $0$ & $-1/2$ & $0$ & $0$ 
& $\U{2+}, \U{2-}$ \\
\hline 
\rom{5}$_{\fs}$ & $0$ & $-1/4$ & $-1/4$ & $-1/4$ 
& $\U{3}, (\U{2+}+\U{2-})$ \\
\hline 
\rom{6}$_{\fs}$ & $0$ & $-1/4$ & $1/4$ & $-1/4$ 
& $\U{3}, (\U{2+}-\U{2-})$ \\
\hline 
\rom{7}$_{\fs}$ & $0$ & $-1/2$ & $0$ & $-1/4$ 
& $\U{3},\U{2+}, \U{2-}$ \\
\hline
\end{tabular}
\caption{The non-Gaussian fixed points for the  couplings in the forward scattering channel, in units of $\pi \eps$.
In the four dimensional coupling space \rom{1}$_\fs$ - \rom{3}$_\fs$ are critical,  \rom{4}$_\fs$ - \rom{6}$_\fs$ are bicritical, and \rom{7}$_\fs$ is tricritical.}
\label{tab:FP-ES-1}
\end{table}
There are $2^3$ solutions to  $\dow_\ell \U{i} = 0$, which   correspond to distinct combinations of the fixed points of $\U{i}$.
The non-Gaussian fixed points  are listed in Table \ref{tab:FP-ES-1}.
There are 3 critical, 3 bicritical, and 1 tricritical fixed points in the four dimensional coupling space.
Among the critical fixed points, the FS-$U(1)$ symmetry emerges  at  \rom{1}$_\fs$, due to the vanishing of $\U{2-}$.
Since it is protected by an emergent symmetry, in the rest of the section we focus on the $\U{2-} = 0$ subspace.
Two more interacting fixed points are  present in the subspace, both of which lose a relevant direction due to the projection to the subspace.
Thus, in the $\U{2-} = 0$ subspace there are 2 critical (\rom{1}$_\fs$, \rom{4}$_\fs$) and 1 bicritical (\rom{7}$_\fs$) fixed points.

The critical points are expected to separate the Weyl-loop semi-metal phase from symmetry broken phases, which are realized by tuning a single parameter.
In the $\U{2-} = 0$ subspace the two critical points \rom{1}$_\fs$ and  \rom{4}$_\fs$ are achieved by tuning $\U{3}$ and $\U{2+}$, respectively.
On tuning these coupling beyond their critical values the system is set to flow towards two distinct strong coupling fixed points.
In this section we determine the stable RG flow trajectories that lead to those fixed points, which will help us identify the possible symmetry broken states that can be realized at finite (or strong) coupling.
We first discuss the SCT originating from \rom{1}$_\fs$, followed by \rom{4}$_\fs$.

Since all couplings but $\U{3}$ vanish at \rom{1}$_\fs$, it is easy to see that the SCT must lie along $\U{3} < -\pi \eps/4$.
This trajectory is stable against small perturbations since the Gaussian fixed points of $\U{i\neq 3}$ are stable.
The effective interaction along the SCT,
\begin{align}
L_{\text{\rom{1}}}^{\fs} = \frac{U_3(\what K_1, \what K_2)}{\kap} ~ \lt( \psi^\dag(\mbf K_1) \sig_3 \psi(\mbf K_1)\rt) 
\lt( \psi^\dag(\mbf K_2) \sig_3 \psi(\mbf K_2)\rt),
\end{align}
indicates that particle-hole pairs  $\psi^\dag(\mbf K) \sig_3 \psi(\mbf K)$ are progressively favored as $|U_3(\what K_1, \what K_2)|$ increases.
Condensation of $\psi^\dag(\mbf K) \sig_3 \psi(\mbf K)$ produces a mass term for the fermions, which gaps out the fermionic excitations.
An identical analysis for the SCT originating at \rom{4}$_\fs$ reveals a stable SCT along $\U{2+} < -\pi \eps/2$, and effective interaction on the SCT, 
\begin{widetext}
\begin{align}
L_{\text{\rom{4}}}^{\fs} = \frac{U_{2+}(\what K_1, \what K_2)}{\kap} ~ 
\lt[
\lt( \psi^\dag(\mbf K_1) \sig_1 \psi(\mbf K_1)\rt) 
\lt( \psi^\dag(\mbf K_2) \sig_1 \psi(\mbf K_2)\rt)
+ \lt( \psi^\dag(\mbf K_1) \sig_2 \psi(\mbf K_1)\rt) 
\lt( \psi^\dag(\mbf K_2) \sig_2 \psi(\mbf K_2)\rt)
\rt].
\end{align}
\end{widetext}

\subsection{Symmetry broken states}
\begin{figure}[!]
\centering
\begin{subfigure}{0.35\columnwidth}
\includegraphics[width=0.95\columnwidth]{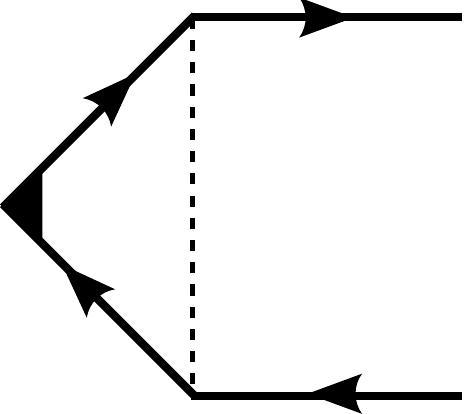}
\caption{}
\label{fig:suscep-DW-a}
\end{subfigure}
\hfill
\begin{subfigure}{0.5\columnwidth}
\includegraphics[width=0.95\columnwidth]{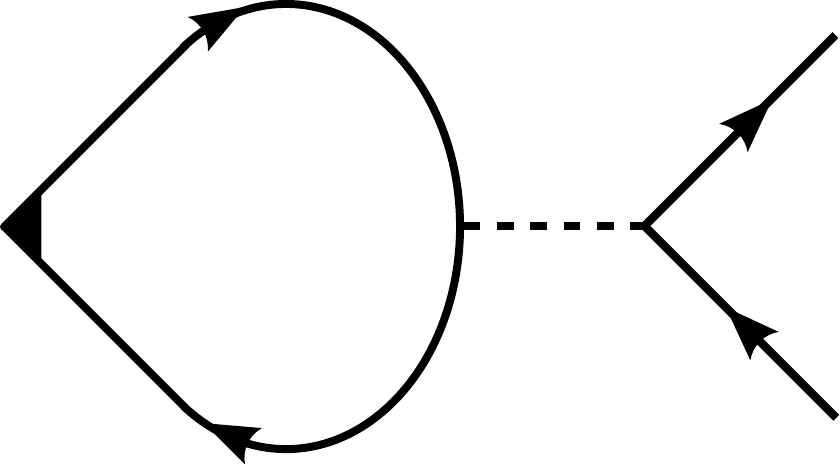}
\caption{}
\label{fig:suscep-DW-b}
\end{subfigure}
\caption{}
\label{fig:suscep-DW}
\end{figure}
In this subsection we compute the anomalous scaling dimension of susceptibility along the two SCTs identified above. 
We also identify the symmetry broken strong coupling fixed points to which the SCTs flow.

Let us consider an insertion of  particle-hole pairs carrying a net momentum $\vec P$ on the plane of the loop,
\begin{align}
S_{2,\mu}^{(DW)}(\vec P) = \int_\Lam \dd{K} \Phi_{\mu}(\vec P; \what K) ~ \bar \psi(P+K) \gam_\mu \psi(K) + h.c.
\end{align}
Here the four-dimensional vector $P \equiv (0, \vec P, 0)$.
In contrast to the pairing susceptibility, the density wave susceptibility obtains quantum correction from the two diagrams in \fig{fig:suscep-DW}.
At low energy quantum corrections to the susceptibility at any finite $\vec P$ are suppressed, compared to $\vec P = 0$.
This is because of a lack of phase space for both the virtual particle and hole to be near the loop when  $\vec P \neq 0$.
Thus we consider the susceptibility for density wave states with $\vec P = 0$.

The source $\Phi_\mu(0;\what K)$ scales as
\begin{align}
& \dow_\ell \ph{0} = \eta \ph{\mu}, \\
& \dow_\ell \ph{1} = \ph{1} \lt[ \eta - \frac{1}{\pi} (\U{2+} +  \U{2-}) \rt], \\
& \dow_\ell \ph{2} = \ph{2} \lt[ \eta - \frac{1}{\pi} (\U{2+} -  \U{2-}) \rt], \\
& \dow_\ell \ph{3} = \ph{3} \lt[ \eta - \frac{2}{\pi} \U{3}  \rt],
\end{align}
where $\eta = 2 -\eps$ is the bare scaling dimension of $\ph{\mu}$.
Thus at the critical point \rom{1}$_\fs$ and the ensuing SCT, only $\ph{3}$ is enhanced, while the scaling dimension of $\ph{\mu \neq 3}$ remain unchanged.
For \rom{4}$_\fs$ and the associated SCT $\ph{1}$ and $\ph{2}$ are equally enhanced.
This degeneracy is protected by the FS-$U(1)$ symmetry. Again the flow equations do not distinguish between angular momentum channel, and the leading instability will be determined by which amgular momentum channel has the largest bare couplings (and $J=0$ is allowed). There is however a constraint, namely that the overall Hamiltonian must be Hermitian. This then enforces that the order parameter must be real i.e. either the instability will be in the $J=0$ channel, or if the instability is in a channel with non-zero angular momentum then a real superposition of $\pm J$ states must arise (i.e. $\propto \sin J \theta$ or $\cos J \theta$). 

The flow out of \rom{1}$_\fs$ is associated with the condensation of $\psi^\dag \sig_3 \psi$. 
If this occurs in a channel with non-zero $J$ then it leads to a low energy Hamiltonian $H \sim (|\vec K| - \kappa) \sigma_1 + K_z \sigma_2 + \Delta_3 \sin(J \theta - \theta_3) \sigma_3$, where $\Dl_3$ and $\theta_3$ are real parameters. 
Such an instability opens a gap almost everywhere on the Weyl loop, with nodes surviving at $\theta = \theta_3 + n \pi/J$ (integer $n$) i.e. this is a gap opening instability that simultaneously breaks the $\theta$-rotational symmetry. 
It also breaks several discrete symmetries, in particular the antiunitary symmetries $\mc P_z \mc P_{\mbf K}$ ($\mc P_\mbf{K}$) for even (odd) $J$, $\mc P_z \mc P_0$, and the mirror symmetry $\mc P_z$. 
However the symmetries $\mc P_{\mbf K}$ ($\mc P_z \mc P_{\mbf K}$) for even (odd) J, and $\mc P_0$ are preserved.  
Meanwhile, if this occurs in a channel with $J=0$ then the gap function is independent of $\theta$, and the $\theta$-rotation  symmetry is preserved, while the discrete symmetries identified above are still broken. 
Condensation in the $J=0$ channel  uniformly gaps out the Weyl loop, with an effective Hamiltonian of the form $H \sim (|\vec K| - \kappa) \sigma_1 + K_z \sigma_2 + \Delta_3 \sigma_3$ (real $\Delta_3$) and a dispersion $E \sim \pm \sqrt{(|\vec K| - \kappa)^2 + K_z^2 + \Delta_3^2}$. Since this is a gap opening instability that preserves an antiunitary symmetry $\mc P_{\mbf K}$, which squares to $-1$, one can ask whether the resulting insulating state is topological or trivial. To address this issue, note that the Weyl loop can be obtained by starting with (spinless) graphene in the $y-z$ plane with it's two (opposite sense) Dirac points located at $\pm \kappa \hat y$ and rotating it through $180$ degrees about the $\hat z$ axis. Gapping out the two Dirac points of (spinless) graphene with a mass term of the same sign on each Dirac point yields a trivial insulator, and rotating a trivial two dimensional insulator through $180$ degrees should yield a trivial three dimensional insulator. 
Nonetheless, we note that on the plane of the Weyl-loop $\mc P_\mbf{K}$ has the interesting  property of mapping the region  outside the loop to its interior, which is an unusual implementation of an anti-unitary symmetry that does not appear to fit naturally into the existing classifications. 


The flow out of critical point \rom{4}$_\fs$ is associated with the condensation of either $\psi^\dag \sig_1 \psi$ or $\psi^\dag \sig_2 \psi$. If this occurs in the $J=0$ channel it leads to an effective Hamiltonian of the form $H \sim (|\vec K| - \kappa) \sigma_1 + K_z \sigma_2 + \Delta_1 \sigma_1 + \Delta_2 \sigma_2$ where $\Delta_{1,2}$ are real parameters. Such a perturbation shifts the radius of the Weyl loop to $\kappa - \Delta$, and shifts it into the plane with $K_z = - \Delta_2$. 
A non-zero $\Delta_1 $ does not break any symmetries and can be absorbed into a redefinition of the Weyl loop radius $\kappa$. 
A non-zero $\Delta_2$ breaks the mirror symmetry, and also the discrete antiunitary symmetries $\mc P_{\mbf K}$ and $\mc P_z \mc P_0$, but preserves $\mc P_z \mc P_{\mbf K}$ and $\mc P_0$ - it simply shifts the Weyl loop out of the $k_z=0$ plane. 
More interesting is the situation where the instability develops in a channel with $J \neq 0$ such that the effective Hamiltonian takes the form $H \sim (|\vec K| - \kappa) \sigma_1 + K_z \sigma_2 + \Delta_1 \sin (J \theta + \theta_1) \sigma_1 + \Delta_2 \cos (J \theta + \theta_2) \sigma_2$ where $\theta_{1,2}$ are constants and $\Delta_{1,2}$ are real. 
Non-zero $\Delta_1$ will lead to a $\theta$-dependent distortion of the nodal ring in the $\vec K$-plane, whereas non-zero $\Delta_2$ will lead to a $\theta$ dependent distortion perpendicular to the x-y plane. 
These order parameters break the $\theta$-rotation symmetry, and correspond to {\it Pomeranchuk} instabilities. The competition between $\Delta_1$ and $\Delta_2$ (in particular whether both $\Delta_1$ and $\Delta_2$ are non-zero, or only one is) will be determined by a Landau Ginzburg calculation similar to those that have already been performed. 
Note that all of these are gapless phases which continue to have a loop of Dirac nodes. 




\section{Conclusion} \label{sec:conclusion}
In this work we analyzed the finite coupling instabilities of a rotationally symmetric Weyl-loop semi-metal in three space dimensions.
The presence of the loop imposes strong kinematic constraints on short-range interactions, similar to those present in a  Fermi liquid. The rotational symmetry of the Weyl loop further endows the problem with enough structure that the functional renormalization group analysis necessary for an extended Fermi surface can be carried out analytically. While the semi-metallic state is stable against weak short-range interactions, symmetry breaking instabilities are present at finite coupling. We deform the dispersion of the system to allow us to access these finite coupling instabilities within the regime of applicability of a weak coupling RG, through an $\epsilon$ expansion type procedure. We find that the only possible instabilities are in the the BCS and the forward scattering channels, which decouple. In the BCS channel the leading instability is to a fully gapped odd angular momentum chiral superconductor, which breaks time reversal symmetry. In the forward scattering channel, various possible instabilities can arise, including a Pomeranchuk instability and a gap opening instability to a trivial insulator. This analysis clarifies what instabilities might be obtained in Weyl loop materials. One question we did not address is the potential competition between instabilities in particle particle and particle hole channels. The Pomeranchuk instabilities in the particle-hole channel can presumably co-exist with superconductivity, whereas the gap opening instability in the particle hole channel is likely to compete with superconductivity. However, a detailed analysis of this interplay is left to future work. 

{\bf Note added}: While finalizing the paper we became aware of a related work \cite{BRoy} that focussed on the Dirac-loop semi-metal using a different regularization scheme than ours. When applied to the Weyl-loop case a subset of our results were obtained

\begin{acknowledgments}
We acknowledge useful conversations with Joseph Maciejko, Sergey Moroz, S.A. Parameswaran, Rahul Roy, S.L. Sondhi, and  Satyanarayan Mukhopadhyay. 
S.S. was supported by the National Science Foundation (Grants No.  DMR-
1004545 and No.  DMR-1442366).
\end{acknowledgments}


\newpage
\appendix
\onecolumngrid
\singlespacing
\section{Computation of quantum corrections} \label{app:dlS}
Here we outline the steps for computation of the one-loop quantum corrections to the quartic vertices.
Since the computation of all the four one-loop vertex corrections follow identical procedure, we provide the details for only the $BCS$ (particle-particle ladder) diagram.
It is useful to list the contraction of various matrix vertices.
Recall that $(\gam_0, \gam_1, \gam_2,\gam_3) \equiv (\sig_3, \sig_2, -\sig_1,I_2)$, therefore $\gamma_1 \gamma_2 = i \gam_0$, $\gam_0 \gam_1 = i \gam_2$, and $\gam_2 \gam_0 = i \gam_1$.
With these results we obtain the multiplication rules for $\gam$-matrices listed in table \ref{tab:mult-gamma}.
\begin{table}[h]
\centering
\begin{subtable}[b]{0.31\textwidth}
\centering
\begin{tabular}{c || c | c | c | c}
\backslashbox{$\gam$}{$\gam'$} & $\gam_3$ & $\gam_0$ & $\gam_1$ & $\gam_2$ \\
\hline \hline
$\gam_3$ & $\gam_0$ & $\gam_3$ & $i \gam_2$ & $- i \gam_1$ \\
\hline
$\gam_0$ & $\gam_3$ & $\gam_0$ & $\gam_1$ & $\gam_2$ \\
\hline
$\gam_1$ & $-i\gam_2$ & $\gam_1$ & $-\gam_0$ & $-i \gam_3$ \\
\hline
$\gam_2$ & $-i\gam_1$ & $\gam_2$ & $i \gam_3$ & $-\gam_0$ \\
\hline
\end{tabular}
\caption{$\gamma \gam_0 \gam'$}
\end{subtable}
\hfill
\begin{subtable}[b]{0.31\textwidth}
\centering
\begin{tabular}{c || c | c | c | c}
\backslashbox{$\gam$}{$\gam'$} & $\gam_3$ & $\gam_0$ & $\gam_1$ & $\gam_2$ \\
\hline \hline
$\gam_3$ & $\gam_1$ & $-i\gam_2$ & $\gam_3$ & $i \gam_0$ \\
\hline
$\gam_0$ & $i\gam_2$ & $-\gam_1$ & $\gam_0$ & $i \gam_3$ \\
\hline
$\gam_1$ & $\gam_3$ & $\gam_0$ & $\gam_1$ & $\gam_2$ \\
\hline
$\gam_2$ & $-i\gam_0$ & $-i \gam_3$ & $\gam_2$ & $-\gam_1$ \\
\hline
\end{tabular}
\caption{$\gamma \gam_1 \gam'$}
\end{subtable}
\hfill
\begin{subtable}[b]{0.31\textwidth}
\centering
\begin{tabular}{c || c | c | c | c}
\backslashbox{$\gam$}{$\gam'$} & $\gam_3$ & $\gam_0$ & $\gam_1$ & $\gam_2$ \\
\hline \hline
$\gam_3$ & $\gam_2$ & $i \gam_1$ & $-i \gam_0$ & $\gam_3$ \\
\hline
$\gam_0$ & $-i\gam_1$ & $-\gam_2$ & $- i \gam_3$ & $\gam_0$ \\
\hline
$\gam_1$ & $i\gam_0$ & $i \gam_3$ & $-\gam_2$ & $\gam_1$ \\
\hline
$\gam_2$ & $\gam_3$ & $\gam_0$ & $\gam_1$ & $\gam_2$ \\
\hline
\end{tabular}
\caption{$\gamma \gam_2 \gam'$}
\end{subtable}
\caption{}
\label{tab:mult-gamma}
\end{table}


\subsection{$BCS$ diagram} \label{app:dlS-BCS}
Contraction of two vertices in the $BCS$ channel leads to the quantum correction,
\begin{align}
\dl S^{(\mu, \nu)}_{int} \Bigr|_{BCS} &=
-\frac{4}{2 \kap^2}  \int \prod_{n=1}^{4} dK_n ~ dK'_n ~  \dl^{(4)}(K_1 - K_2 + K_3 - K_4) ~  \dl^{(4)}(K'_1 - K'_2 + K'_3 - K'_4) \nn \\
& \quad \times  \gam_\mu^{a_1,b_1} \gam_\mu^{a_2, b_2} \gam_\nu^{a'_1,b'_1} \gam_\nu^{a'_2, b'_2} ~ u_{\mu}(\what K_1, \what K_2,\what K_3,\what K_4) ~ u_{\nu}(\what K'_1, \what K'_2,\what K'_3,\what K'_4) \nn \\
& \quad \times  \avg{\psi_{b'_2}(K'_4) \bar \psi_{a_1}(K_1)} \avg{\psi_{b'_1}(K'_2) \bar \psi_{a_2}(K_3)}  \bar \psi_{a'_2}(K'_3)  \psi_{b_1}(K_2) \bar\psi_{a'_1}(K'_1)  \psi_{b_2}(K_4).
\label{eq: delS-1}
\end{align}
Here for notational and computational convenience  we have used $u_\mu$ to identify the coupling functions for the $(\bar \psi ~ \gam_\mu ~ \psi)^2$ vertex.
In particular
\begin{align}
u_0 = g_1, \quad u_1 = -(g_{2+} + g_{2-}), \quad u_2 = -(g_{2+} - g_{2-}), \quad u_3 = g_3, 
\label{eq:u-g}
\end{align}
where we have suppressed the dependence on loop-coordinates on both sides.
Utilizing the definition of the propagator in \eq{eq: propagator}, and integrating over $K'_2$, $K'_4$, and $K_3$ leads to
\begin{align}
\dl S^{(\mu, \nu)}_{int} \Bigr|_{BCS} &=
-\frac{2}{\kap^2}  \int dK_2 dK_4 dK'_1 dK'_3 ~ (2\pi)^{4} \dl^{(4)}(K'_3 - K_2 + K'_1 - K_4) \nn \\
& \qquad \times  \Upsilon_{\mu  \nu}^{a'_2,b_1; a'_1,b_2}(K'_3, K_2, K'_1, K_4) ~~ \bar{ \psi}_{a'_2}(K'_3)  \psi_{b_1}(K_2) \bar\psi_{a'_1}(K'_1)  \psi_{b_2}(K_4),
\label{eq: delS-2}
\end{align}
where
\begin{align}
\Upsilon_{\mu \nu}(K_1,K_2,K_3, K_4) &= \int dQ ~ \Bigl[ \gam_\nu ~G(Q)~ \gam_\mu \Bigr] \circ \Bigl[ \gam_\nu ~G(K_4 + K_2 - Q)~ \gam_\mu \Bigr] \nn \\ 
& \times  u_{\mu}(\what Q, \what K_4,\what{(K_2 + K_4 - Q)},\what K_2) ~ u_{\nu}(\what K_1, \what{(K_2 + K_4 - Q)},\what K_3,\what Q),
\label{eq: ups-1}
\end{align}
with $A \circ B \equiv A_{a_1, b_1}  B_{a_2, b_2}$, and $\what{(K_2 + K_4 - Q)}$ being the unit vector along $(\vec K_2 + \vec K_4 - \vec Q)$.
Here the internal momentum $\mbf Q$ is restricted to lie within the shell being eliminated (c.f. \fig{fig:coarse-grain}).

The BCS channel is defined by $(\what{K}_1, \what{K}_2) = ( -\what{K}_3, - \what{K}_4)$.
Because we are interested in the IR, we set the external frequency to $0$, and external  momenta to lie on the loop.
Combined with the angular constraint, this implies $\vec K_2 + \vec K_4 = 0 = \vec K_1 + \vec K_3$.
Therefore,
\begin{align}
\Upsilon_{\mu \nu}(\what K_1, \what K_2, -\what K_1, -\what K_2) 
&= \int dQ ~ \Bigl[ \gam_\nu ~G(Q)~ \gam_\mu \Bigr] \circ
\Bigl[ \gam_\nu ~G(-Q)~ \gam_\mu \Bigr] \nn \\ 
& \times  u_{\mu}(\what Q, -\what K_2, - \what{Q}, \what K_2) ~ u_{\nu}(\what K_1, - \what{Q}, - \what K_1, \what Q).
\label{eq: ups-BCS-1}
\end{align}
Let us define 
\begin{align}
\Lam^{-\eps} V'_\mu(\theta_1 - \theta_2) = u_\mu(\vu K_1, \vu K_2, -\vu K_1, -\vu K_2),
\end{align}
where $\theta_i$ is the angle the loop momentum $\kap \what K_i$ makes with respect to the $x$-axis.
Thus, we obtain an equivalent expression to \eq{eq: ups-BCS-1} in terms of the angles, 
\begin{align}
\Upsilon_{\mu \nu}(\theta_1, \theta_2, \theta_1 + \pi, \theta_2 + \pi) 
&=  
\kap \Lam^{-2\eps} \int_{-\pi}^{\pi} \frac{d\theta d\vphi}{(2\pi)^2} \int_{-\infty}^{\infty} \frac{dq_0}{2\pi} \int_{(1-d\ell)\Lam}^{\Lam} \frac{dq ~ q}{2\pi} 
 ~ V'_{\mu}(\theta - \theta_2 +\pi) V'_{\nu}(\theta_1 - \theta -\pi) \nn \\
& \quad \times  
\Bigl[ \gam_\nu ~G(q_0, q, \vphi)~ \gam_\mu \Bigr] \circ  \Bigl[ \gam_\nu ~G(- q_0, q, -\vphi)~ \gam_\mu \Bigr].
\label{eq: ups-2}
\end{align}
We note that $G(q_0, q, \vphi)$ does not have a specific parity under spacetime inversion.
As a result, the integrand of \eq{eq: ups-2}, up to terms that are even in $\vphi$ equals,
\begin{align}
\frac{q_0^2 [\gam_\nu \gam_0 \gam_\mu] \circ [\gam_\nu \gam_0 \gam_\mu] 
- q^{2(2-\eps)}(\cos^2{\vphi} ~ [\gam_\nu \gam_1 \gam_\mu] \circ [\gam_\nu \gam_1 \gam_\mu] 
- \sin^2{\vphi} ~ [\gam_\nu \gam_2 \gam_\mu] \circ [\gam_\nu \gam_2 \gam_\mu] ) }
{(q_0^2 + q^{2(2-\eps)})^2},
\label{eq: BCS-integrand}
\end{align}
The opposite sign for the $\gam_1$ and $\gam_2$ terms will lead to unequal quantum corrections to the $\gam_1$ and $\gam_2$ vertices, as we will see below.
This is a manifestation of the absence of $\mc R_\vphi$ symmetry for the interaction vertices, in general.

Noting that $\theta$ decouples from rest of the internal variables, it is simplest to integrate in the order $\vphi$, $q$, and $q_0$.
We cannot explicitly integrate over $\theta$, but we can simplify the $V'_\mu$ dependence of the quantum correction by expressing the coupling functions in terms of angular momentum harmonics. 
The inverse Fourier transform of $V'_\mu(\theta)$ is given by
\begin{align}
V'_\mu(\theta) = \sum_{J = -\infty}^{\infty} ~ e^{-i\theta  J} ~ \wV{\mu},
\end{align}
which leads to
\begin{align}
\int_{-\pi}^{\pi} \frac{d\theta}{2\pi} ~ V'_{\mu}(\theta - \theta_2 +\pi) V'_{\nu}(\theta_1 - \theta -\pi)
& = \sum_J e^{-i(\theta_1 - \theta_2) J} ~ \wV{\mu} \wV{\nu}
\end{align}
Therefore, \eq{eq: ups-2} evaluates to,
\begin{align}
& \Upsilon_{\mu \nu}(\theta_1, \theta_2, \theta_1 + \pi, \theta_2 + \pi) 
= 
- d\ell ~\frac{\kap \Lam^{-\eps}}{16\pi}  \sum_J e^{-i(\theta_1 - \theta_2) J} ~ \wV{\mu} \wV{\nu} \nn \\
& \qquad \times 
\lt[ 
(\gam_\nu \gam_1 \gam_\mu) \circ (\gam_\nu \gam_1 \gam_\mu) 
- (\gam_\nu \gam_2 \gam_\mu) \circ (\gam_\nu \gam_2 \gam_\mu) 
- 2(\gam_\nu \gam_0 \gam_\mu) \circ (\gam_\nu \gam_0 \gam_\mu) \rt].
\end{align}
This leads to quantum corrections to the BCS channels,
\begin{align}
& \dl S^{(\mu, \nu)}_{int} \Bigr|_{BCS} =
 \frac{\Lam^{-\eps} d\ell}{8\pi \kap} \sum_J \int \prod_{n=1}^{4} dK_n ~ (2\pi)^{4} \dl^{(4)}(K_1 - K_2 + K_3 - K_4) ~ \dl(\mbf K_1 + \mbf K_3) ~ \dl(\mbf K_2 + \mbf K_4) \nn \\
& \qquad  
\times e^{-i(\theta_1 - \theta_2)J} ~\wV{\mu} \wV{\nu} \Bigl[ 
\lt\{\bar{ \psi}(K_1) ~\gam_\mu  \gam_1 \gam_\nu ~ \psi(K_2) \rt\} 
 \lt\{\bar\psi(K_3) ~ \gam_\mu  \gam_1 \gam_\nu ~ \psi(K_4) \rt\} \nn \\
&
\quad  - \lt\{\bar{ \psi}(K_1) ~\gam_\mu  \gam_2 \gam_\nu ~ \psi(K_2) \rt\} 
 \lt\{\bar\psi(K_3) ~ \gam_\mu  \gam_2 \gam_\nu ~ \psi(K_4) \rt\}
- 2\lt\{\bar{ \psi}(K_1) ~\gam_\mu  \gam_0 \gam_\nu ~ \psi(K_2) \rt\} 
 \lt\{\bar\psi(K_3) ~ \gam_\mu  \gam_0 \gam_\nu ~ \psi(K_4) \rt\} \Bigr].
\label{eq: delS-3}
\end{align}
The net quantum correction is obtained by summing over $\mu$ and $\nu$,
\begin{align}
& \dl S_{int}^\bcs = \sum_{\mu, \nu = 0}^{3} \dl S^{(\mu, \nu)}_{int} \Bigr|_{BCS} \nn \\
&=
-\frac{\Lam^{-\eps} d\ell}{4\pi \kap} \sum_J  \int \prod_{n=1}^{4} dK_n ~ (2\pi)^{4} \dl^{(4)}(K_1 - K_2 + K_3 - K_4) ~ \dl(\mbf K_1 + \mbf K_3) ~ \dl(\mbf K_2 + \mbf K_4) ~ e^{-i(\theta_1 - \theta_2)J}\nn \\
&  \times  \Bigl[
\lt\{ 2(\wV{3} \wV{0} - \wV{1} \wV{2} ) + (\wV{0} \wV{2} + \wV{3} \wV{2} 
- \wV{3} \wV{1} - \wV{0} \wV{1} ) \rt\} ~ (\bar{ \psi} ~ \psi)^2 \nn \\
&\quad + \lt\{ ( (\wV{3})^2 + (\wV{0})^2 + (\wV{1})^2 + (\wV{2})^2 ) 
+ (\wV{0} \wV{2} + \wV{3} \wV{2} - \wV{0} \wV{1} - \wV{3} \wV{1} ) \rt\} 
~(\bar{ \psi} \gam_0 \psi)^2 \nn \\
& \quad + \lt\{ 2(\wV{0} \wV{1} - \wV{3} \wV{2}) 
+ (\wV{1} \wV{2} - \wV{3} \wV{0} ) 
- \half( (\wV{3})^2 + (\wV{0})^2 + (\wV{1})^2 + (\wV{2})^2 ) \rt\} (\bar \psi \gam_1 \psi)^2\nn \\
&\quad  + \lt\{ 2(\wV{0} \wV{2} - \wV{3} \wV{1}) 
+ (\wV{3} \wV{0} - \wV{1} \wV{2} )
 + \half( (\wV{3})^2 + (\wV{0})^2 + (\wV{1})^2 + (\wV{2})^2 ) \rt\} (\bar \psi \gam_2 \psi)^2 \Bigr],
\label{eq:delS-4}
\end{align}
where the dependence of $\psi$ on $K_n$ is made implicit for notational convenience.

\subsection{Non-$BCS$ diagrams} \label{app:dlS-FS}
The forward scattering channels are renormalized by Figs. \ref{fig:PH}, \ref{fig:bubble}, and \ref{fig:penguin}.
In order to compute their contributions it is convenient to distinguish between the FS and ES channels at an intermediate step, and unify them at the end through the relationship
\begin{align}
\begin{pmatrix}
 {U'_{0}}^{\es} \\  {U'_{1}}^{\es} \\ {U'_{2}}^{\es} \\ {U'_{3}}^{\es} 
\end{pmatrix}
= \frac{1}{2} \left(
\begin{array}{cccc}
 -1 & -1 & 1 & 1 \\
 -1 & 1 & -1 & 1 \\
 -1 & 1 & 1 & -1 \\ 
 -1 & -1 & -1 & -1 \\
\end{array}
\right)
\begin{pmatrix}
 {U'_{0}}^{\fs} \\  {U'_{1}}^{\fs} \\ {U'_{2}}^{\fs} \\ {U'_{3}}^{\fs} 
\end{pmatrix},
\label{eq:ES-FS}
\end{align}
where the primed and their unprimed counterparts defined in the main section are related in the same way as \eq{eq:u-g}.
The relationship between the ES and FS representation of the coupling functions in the forward scattering channel is defined on the loop through the equivalence of $L_{int}^{\es} \equiv  \sum_\mu {U'_{\mu}}^{\es} (\bar \psi \gam_\mu \psi)^2_{\es} = \sum_\mu {U'_{\mu}}^{\fs} (\bar \psi \gam_\mu \psi)^2_{\fs} \equiv L_{int}^{\fs},$ where the subscript in the $\psi^4$ term denotes the arrangement of the fermion momenta in accordance with the definition of the FS and ES channels.

At leading order in $\Lam/\kap$ the external legs of the $ZS'$  diagram are arranged as in the ES channel, while those of the $ZS$ and $P$ diagrams are arranged as  in the FS channel.
Repeating the computation presented above for the $BCS$ diagrams to the present set of diagrams leads to the quantum corrections
\begin{align}
\dl S_{int}^{(ZS')} &= 
\frac{\Lam^{-\eps} d\ell}{4\pi \kap} \sum_J  \int \prod_{n=1}^{4} dK_n ~ (2\pi)^{4} \dl^{(4)}(K_1 - K_2 + K_3 - K_4) ~ \dl(\mbf K_1 - \mbf K_4) ~ \dl(\mbf K_3 - \mbf K_2) ~ e^{-i(\theta_1 - \theta_2)J}\nn \\
&  \times  \Bigl[
\lt\{ 2(\wUe{0}  \wUe{3} + \wUe{1}  \wUe{2}) + \wUe{1}  \wUe{3} + \wUe{0}  \wUe{2} + \wUe{2}  \wUe{3} + \wUe{0}  \wUe{1} \rt\} ~ (\bar{ \psi} ~ \psi)^2 \nn \\
&\quad + \lt\{ \sum_{\mu} (\wUe{\mu})^2 + (\wUe{0} + \wUe{3}) (\wUe{1} + \wUe{2}) \rt\} 
~(\bar{ \psi} \gam_0 \psi)^2 \nn \\
& \quad + \lt\{ \half\sum_{\mu} (\wUe{\mu})^2 + 2(\wUe{0}  \wUe{1} + \wUe{2} \wUe{3}) + (\wUe{0}  \wUe{3} + \wUe{1} \wUe{2}) \rt\} (\bar \psi \gam_1 \psi)^2\nn \\
&\quad  + \lt\{ \half\sum_{\mu} (\wUe{\mu})^2 + 2(\wUe{0}  \wUe{2} + \wUe{1} \wUe{3}) + (\wUe{0}  \wUe{3} + \wUe{1} \wUe{2}) \rt\} (\bar \psi \gam_2 \psi)^2 \Bigr],
\label{eq:delS-ZSp} \\
\dl S_{int}^{(ZS)} &= 
-\frac{\Lam^{-\eps} d\ell}{2\pi \kap} \sum_J  \int \prod_{n=1}^{4} dK_n ~ (2\pi)^{4} \dl^{(4)}(K_1 - K_2 + K_3 - K_4) ~ \dl(\mbf K_1 - \mbf K_2) ~ \dl(\mbf K_3 - \mbf K_4) ~ e^{-i(\theta_1 - \theta_2)J}\nn \\
&  \times  \Bigl[ 
2(\wUf{3})^2 ~ (\bar{ \psi} ~ \psi)^2 
- (\wUf{1})^2 ~ (\bar{ \psi} \gam_1 \psi)^2 
- (\wUf{2})^2 ~ (\bar{ \psi} \gam_2 \psi)^2 
\Bigr]
\label{eq:delS-ZS} \\
\dl S_{int}^{(P)} &= 
\frac{\Lam^{-\eps} d\ell}{2\pi \kap} \sum_J  \int \prod_{n=1}^{4} dK_n ~ (2\pi)^{4} \dl^{(4)}(K_1 - K_2 + K_3 - K_4) ~ \dl(\mbf K_1 - \mbf K_2) ~ \dl(\mbf K_3 - \mbf K_4) ~ e^{-i(\theta_1 - \theta_2)J}\nn \\
&  \times  \Bigl[ 
2\wUf{3} (\sum_{\mu} \wUe{\mu}) ~ (\bar{ \psi} ~ \psi)^2 
- \wUf{1}(\wUe{3} - \wUe{0} + \wUe{1} - \wUe{2}) ~ (\bar{ \psi} \gam_1 \psi)^2 \nn \\
& \qquad - \wUf{2}(\wUe{3} - \wUe{0} - \wUe{1} + \wUe{2}) ~ (\bar{ \psi} \gam_2 \psi)^2 
\Bigr]
\label{eq:delS-P}
\end{align}
Applying the transformation in \eq{eq:ES-FS} leads to the net quantum correction to the forward scattering vertices in the FS representation (we drop explicit reference to FS),
\begin{align}
\dl S_{int}^{\fs} &= 
-\frac{2\Lam^{-\eps} d\ell}{\pi \kap} \sum_J  \int \prod_{n=1}^{4} dK_n ~ (2\pi)^{4} \dl^{(4)}(K_1 - K_2 + K_3 - K_4) ~ \dl(\mbf K_1 - \mbf K_2) ~ \dl(\mbf K_3 - \mbf K_4) ~ e^{-i(\theta_1 - \theta_2)J}\nn \\
&  \times  \Bigl[ 
2(\wU{3})^2 ~ (\bar{ \psi} ~ \psi)^2 
- (\wU{1})^2 ~ (\bar{ \psi} \gam_1 \psi)^2 
- (\wU{2})^2 ~ (\bar{ \psi} \gam_2 \psi)^2 
\Bigr].
\label{eq:delS-FS}
\end{align}


\section{Susceptibilities} \label{app:chi}
In this section we outline the computation of the anomalous dimension of the susceptibilities of both pairing and density-wave channels. 

\subsection{Pairing susceptibility} \label{app:BCS-chi}
Let the quadratic-insertion be
\begin{align}
S_{2,\mu}^{\bcs} = \int_{\Lam} dK ~ \conj{\Dl_\mu}(\what K) \trans{\psi}(-K) \gam_\mu \psi(K) + h.c. 
\label{eq:insert-BCS}
\end{align}
The label $\Lam$ reminds us of the cutoff for the effective action.
The quantum corrections to \eq{eq:insert-BCS} are generated by contracting it with the quartic vertices in the action, which results in processes represented by \fig{fig:suscep}, 
\begin{align}
\dl S_{2,\mu}^{\bcs} = -\sum_\nu \int_{(1 - d\ell)\Lam} \dd{K}  \trans{\psi}(-K) ~\trans{\gam_\nu}  \Gam_{\bcs}^{\mu \nu}(\hat K) \gam_\nu ~ \psi(K) + h.c., 
\end{align}
where 
\begin{align}
\Gam_{\bcs}^{\mu \nu}(\hat K) 
= \frac{\Lam^{-\eps}}{\kap} \int' \dd{Q}  V'_\nu(\hat Q, \hat K) \conj{\Dl_\mu}(\hat Q) ~ \trans{G}(-Q) \gam_\mu G(Q).
\end{align}
Here the primed integration sign implies that the integral is restricted within the high-energy region which corresponds to $k \in [(1- d\ell) \Lam, \Lam]$.
After introducing the angular momentum harmonics, the mode elimination leads to
\begin{align}
\Gam_{\bcs}^{\mu \nu}(\theta_k)  = \dd{\ell}  \frac{1}{8\pi} 
\lt[ \gam_0 \gam_\mu \gam_0 + \half \sum_{i=1,2} \gam_i \gam_\mu \gam_i  \rt]
\sideset{}{'}\sum_J ~ e^{-i l \theta_k} ~ \wV{\nu} \conj{\D{\mu}},
\end{align}
where the prime over the sum represent the restriction of $J$ to  even (odd) integers when  $\trans{\gam_\mu} = \gam_\mu$ ($\trans{\gam_\mu} = - \gam_\mu$).
Substituting $\Gam_{\bcs}^{\mu \nu}$ into the expression of quantum correction, and summing over $\nu$  results in
\begin{align}
\dl S_{2,\mu}^{\bcs} 
=  \dd{\ell}    
\sideset{}{'}\sum_J \int_{(1-d\ell) \Lam} \dd{K} \dl_\mu(\{\V{i}\}) ~ \conj{\D{\mu}} ~ \trans{\psi}(-K) ~ \gam_\mu ~ \psi(K) + h.c., 
\end{align}
where
\begin{align}
& 8\pi ~ \dl_0(\{\V{i}\}) = 0, \nn \\
& 8\pi ~\dl_1(\{\V{i}\}) = - (\V{1} - \V{3}) + 2 \V{2+} = 2 \f{1-}, \nn \\
& 8\pi ~\dl_2(\{\V{i}\}) = - (\V{1} - \V{3}) - 2 \V{2+} = - 2 \f{1+}, \nn \\
& 8\pi ~\dl_3(\{\V{i}\}) = - 2(\V{1} + \V{3}) - 4\V{2-} = -4\f{2+}.
\label{eq:dl-BCS}
\end{align}
Adding the quantum correction to $\dl S_{2,\mu}^{\bcs}$ and rescaling all dimensionful quantities to restore the cutoff to $\Lam$ gives the beta function of $\Dl_{\mu,l}$,
\begin{align}
\dow_\ell \D{\mu} = \D{\mu} \lt[ \eta +  \dl_\mu(\{\V{i}\})\rt].
\end{align}


\subsection{Density wave  susceptibility} \label{app:DW-chi}
Figs. \ref{fig:suscep-DW-a} and \ref{fig:suscep-DW-b} leads to the quantum corrections, 
\begin{align}
& \dl S_{2,\mu}^{\dw}(\vec P) \Bigr|_a = -2 \sum_\nu \int_{\Lam(1-d\ell)} \dd{K}  \bar \psi(P+K) ~ \gam_\nu \Gam_{\dw;a}^{\mu\nu}(\what P, \what K) \gam_\nu ~ \psi(K) + h.c., \nn \\
&\dl S_{2,\mu}^{\dw}(\vec P) \Bigr|_b =  2 \sum_\nu \int_{\Lam(1-d\ell)} \dd{K} \Gam_{\dw;b}^{\mu\nu}(\what P, \what K) ~ \bar \psi(P+K) ~ \gam_\nu ~ \psi(K) + h.c., 
\label{eq:DW-1}
\end{align}
respectively, where ($u_\mu$ are defined in \eq{eq:u-g})
\begin{align}
&\Gam_{\dw;a}^{\mu\nu}(\what P, \what K) 
= \frac{1}{\kap} \int' \dd{Q}  \Phi_{\mu}(\vec P; \what Q) ~ u_\nu(\what Q, \what K, \what{(K+P)}, \what{(Q + P)}) ~ G(Q+P) \gam_\mu G(Q), \nn \\
& \Gam_{\dw;b}^{\mu\nu}(\what P, \what K) 
= \frac{1}{\kap} \int' \dd{Q}  \Phi_{\mu}(\vec P; \what Q) ~ u_\nu(\what Q, \what {(Q + P)}, \what{(K+P)}, \what{K}) ~ \tr{\gam_\nu G(Q+P) \gam_\mu G(Q)}.
\label{eq:DW-2}
\end{align}
The difference in overall sign between the two quantum corrections in \eq{eq:DW-1} arises from the fermion-loop in \fig{fig:suscep-DW-b}.
We set $\vec P = 0$, and apply the steps in appendices \ref{app:dlS-FS} and \ref{app:BCS-chi} to obtain the beta functions for $\Phi_{\mu, J}$ quoted in the main section.


\begin{thebibliography}{99}
\bibitem{nonlocality} Tuning $\eta$ away from integer values introduces non-locality in the fermion dispersion,  because dispersions with non-integer  $\eta$ cannot be represented by local operators in coordinate space. Since perturbative RG generates only local terms, notwithstanding logarithms, non-local vertices remain unrenormalized.
Although this artifact affects the  estimates for  a subset of critical exponents, one-loop vertex corrections are not expected to be affected.
\end{thebibliography}
\end{document}